\documentclass[trackchanges]{aastex701}

\usepackage{multirow}

\begin{document}

\title{Metallicity regulates planet formation across all masses}

\author[sname='Nguyen']{Max Nguyen}
\affiliation{Leland High School, 6677 Camden Avenue, San Jose, CA 95120, USA.}
\email[show]{MaxHungNguyen@gmail.com}  

\author[orcid=0000-0002-0601-6199,gname=Vardan, sname='Adibekyan']{Vardan Adibekyan} 
\affiliation{Instituto de Astrof\'isica e Ci\^encias do Espa\c{c}o, Universidade do Porto, CAUP,  \\ \small Rua das Estrelas, Porto, 4150-762 , Portugal}
\email{vadibekyan@astro.up.pt}


\begin{abstract}

The role of stellar metallicity in shaping planetary systems is central to our understanding of planet formation. While the core accretion paradigm is widely accepted as the dominant mechanism for forming low- and intermediate-mass planets, the origin of the most massive planets remains debated, with gravitational instability often invoked to explain their existence. In this study, we analyze the dependence of planet formation on metallicity using the total heavy-element mass fraction ($Z$), which is a proxy for the composition of the protoplanetary disk inferred from stellar photospheres. We show that even the most massive planets form preferentially in metal-rich environments. $Z$ correlates not only with the presence of planets, but also with planetary system multiplicity and total planetary mass. The most massive planets are found in the most metal-rich environments, and, in agreement with core-accretion theory, only the upper end of the planetary mass distribution shows a clear positive correlation with metallicity. These findings suggest that the chemical enrichment of protoplanetary disks plays a central role in shaping the full spectrum of planetary masses.

\end{abstract}



\section{Introduction} 

Exoplanet research is fundamentally intertwined with stellar astrophysics. The key physical properties of exoplanets, such as radius and mass, are typically derived relative to their host stars \citep{Adibekyan-18}. Moreover, because protoplanetary disks dissipate within a few million years, most known exoplanets are no longer embedded in their original formation environments. As stars and planets originate from the same molecular cloud material, stellar atmospheres serve as the primary fossil record of the conditions present during planet formation \citep{Grossman-72, Adibekyan-21}.

A powerful approach for understanding how planets form and evolve is to investigate correlations between planetary occurrence rates and properties and the characteristics of their host stars \citep{VanEylen-18, Mulders-18, Adibekyan-21, Banerjee-24, Weeks-25}. In particular, stellar metallicity has proven valuable as a tracer of planet formation efficiency and architecture \citep{Santos-04, Fischer-05, Sousa-11, Buchhave-12, Adibekyan-13, Buchhave-14}. Observations show that the occurrence rate of low-mass planets appears largely independent of stellar metallicity \citep{Sousa-11, Buchhave-12, Kutra-21}. However, this relation is not uniform for hot and warm planets, a positive correlation with metallicity has been observed for hot low-mass planets \citep[e.g.,][]{Mulders-16, Petigura-18, Wilson-22, Zink-23}. In contrast, Jupiter-mass planets are much more common around metal-rich stars \citep{Santos-04, Fischer-05}.

Overall, these trends are consistent with predictions from the core-accretion (CA) model, where higher metallicity promotes the rapid formation of solid cores necessary for gas giant formation. However, the formation of the most massive planets, super-Jupiters with masses greater than about 4 $M_{\mathrm{J}}$, remains uncertain. Some studies have reported that these planets tend to orbit stars with lower metallicities than those hosting Jupiter-mass planets \citep{Santos-17, Maldonado-19, Matsukoba-23}, suggesting that gravitational instability (GI) may dominate in that mass regime.

Most previous studies have relied on stellar iron abundance ([Fe/H]) as a proxy for the total metallicity of the protoplanetary disk. While this approximation may be adequate for near-solar metallicities, it becomes increasingly inaccurate at lower metallicities, where significant $\alpha$-element enhancement is observed \citep{Adibekyan-12, Ratcliffe-23}. Furthermore, different metals contribute differently to planet formation. Thus, a physically more meaningful metric is the summed mass fraction of heavy elements ($Z$), which captures the abundance of elements that are most relevant to planet formation \citep{Santos-17b}.

Recently, \citet{Nguyen-24} using the heavy-element fraction $Z$ instead of [Fe/H] showed that super-Jupiters form in environments as metal-rich as those hosting Jupiter-mass planets, contradicting earlier claims and potentially lending support to the CA scenario. This is particularly relevant given that the role of metallicity in GI models remains poorly understood \citep{Helled-14, Vorobyov-25}.

In this study, we investigate how the presence of planets in different mass regimes depends on the summed mass fraction of heavy elements in planet-building blocks, inferred from host star atmospheric abundances. Additionally, we examine whether $Z$ influences planetary system multiplicity, total planetary mass, and the mass of the heaviest planet in each system. Ultimately, our analysis aims to clarify the role of metallicity in shaping the formation and architecture of planetary systems across the full spectrum of planetary masses, from Earth-mass planets to brown dwarfs.

\section{Sample selection}\label{sec:sample}

We began our sample selection from the NASA Exoplanet Archive \citep[NEA,][]{nea}.  At the time of retrieval\footnote{The data were retrieved on 18/12/2024}, the NEA listed 5,806 confirmed exoplanets orbiting 4,069 stars. Of these planets, 2,593 had measured masses below the brown dwarf limit of 13\,$M_{\mathrm{J}}$ \citep{Spiegel-11}, and about 84\% of these mass measurements had relative uncertainties below 50\%. We used the parameter \texttt{pl\_bmassj} from the NEA, which provides the best available planet mass estimates, prioritizing direct mass measurements, $M \sin(i)/\sin(i)$, or $M \sin(i)$, depending on availability. However, many of these mass estimates were derived indirectly from mass-radius relationships. To ensure reliable and uniform mass measurements, we restricted our analysis to planets detected via the radial velocity (RV) method and with relative mass uncertainties below 50\%. This criterion yielded a sample of 959 planets orbiting 715 host stars (about 97\% of all RV-detected planets).

Next, we cross-matched the planet-hosting stars with the Hypatia Catalog \citep{Hinkel-14} to retrieve stellar elemental abundances. The Hypatia Catalog is a carefully curated compilation of abundance measurements for nearby stars within 500 pc of the Sun, as well as all known exoplanet host stars regardless of distance, although it is not fully homogeneous. We first restricted the sample to stars with elemental abundances $[X/H] < 0.6$ dex, since the Galaxy has not enriched beyond this metallicity threshold \citep{Ratcliffe-23}. After this selection, we identified 463 host stars (harboring 653 planets) with available abundance measurements for carbon (C), oxygen (O), magnesium (Mg), silicon (Si), and iron (Fe). To ensure high data quality, we further limited the sample to stars with effective temperatures ($T_{\mathrm{eff}}$) between 4500 and 6500 K, which yielded 429 stars hosting 605 planets. Finally, rather than imposing constraints on the precision of individual abundances, and since our focus is on the heavy-element fraction $Z$, we restricted the sample to stars with a relative uncertainty in $Z$ below 50\%. The resulting final sample comprises 412 stars hosting 575 planets, with relative mass uncertainties averaging 12$\pm$8\% (median 9$\pm$6\%). The mean relative uncertainty in $Z$ for this sample is 17\%, with a median of 13\%.

To explore how heavy-element content relates to planet mass and system architecture, we divided our sample into four planetary mass regimes: sub-Neptunes (0–0.05 $M_{\mathrm{J}}$), Neptunes (0.05–0.15 $M_{\mathrm{J}}$), Jupiters (0.3–4 $M_{\mathrm{J}}$), and super-Jupiters (4–13 $M_{\mathrm{J}}$). It should be noted that there is no universally adopted set of mass boundaries for planetary populations. Our limits reflect representative values frequently used in the literature and capture the broad distinctions between the main planet classes. The intermediate regime between 0.15 and 0.30 $M_{\mathrm{J}}$, often referred to as the “planetary desert” \citep[e.g.,][]{Ida-04}, was excluded because such planets cannot be unambiguously classified as Neptunes (nor Jupiters) and their relatively small numbers mean that their inclusion would not affect the statistical results for the Jupiter population.

We applied two selection strategies: classifying planetary systems based on their most massive planet, and classifying based on the presence of any planet in each mass range. Table~\ref{tab:planet_counts} summarizes the number of planets and host stars in each category for both strategies.

\begin{table}[h]
\centering
\caption{Number of planets and host stars in each mass regime under two selection strategies: based on the most massive planet in the system and based on the presence of any planet in the mass range.}
\label{tab:planet_counts}
\begin{tabular}{lcc}
\hline
\textbf{Mass regime} & \textbf{Planets / Hosts (most massive)} & \textbf{Planets / Hosts (any)} \\
\hline
sub-Neptunes ($M_{\mathrm{p}}$ $\leq$ 0.05 $M_{\mathrm{J}}$)       & 70 / 36  & 97 / 55 \\
Neptunes (0.05 $<$ $M_{\mathrm{p}}$  $\leq$ 0.15  $M_{\mathrm{J}}$) & 44 / 28  & 48 / 40 \\
Jupiters (0.3 $<$ $M_{\mathrm{p}}$  $\leq$ 4  $M_{\mathrm{J}}$)     & 298 / 228 & 294 / 248 \\
Super-Jupiters (4 $<$ $M_{\mathrm{p}}$  $\leq$ 13  $M_{\mathrm{J}}$) & 132 / 96 & 101 / 96 \\
\hline
\end{tabular}
\end{table}

As the main control sample, we selected 4,688 stars without known planets from the Hypatia Catalog, requiring available abundance measurements for C, O, Mg, Si, and Fe within the range of –1 to 0.6 dex, consistent with the abundances of the planet-hosting stars. We further restricted the sample to stars with effective temperatures $T_{\mathrm{eff}}$ between 4500 and 6500 K and surface gravities $2 \leq \log g \leq 5$ dex, corresponding to the parameter space occupied by the host-star sample. Finally, we excluded stars with relative uncertainties in $Z$ larger than 50\%. This resulted in a control sample of 3,673 stars, with a mean relative uncertainty in $Z$ of 14\% and a median of 12\%.

Since these stars have not necessarily been monitored for planets, we also constructed a second control sample consisting of 431 non-host stars from the HARPS planet-search program \citep{Mayor-03, Santos-11}, for which abundances of the same elements were homogeneously determined \citep{Delgado-21}. Applying the same restrictions as for the Hypatia and host-star samples yielded a final HARPS non-host sample of 394 stars, with a mean relative uncertainty in $Z$ of 20\% and a median of 18\%. Although smaller in size, this sample is particularly valuable because these stars have been systematically monitored for planets, with none detected to date.

\begin{figure*}[h]
    \centering
    \includegraphics[width=0.6\textwidth]{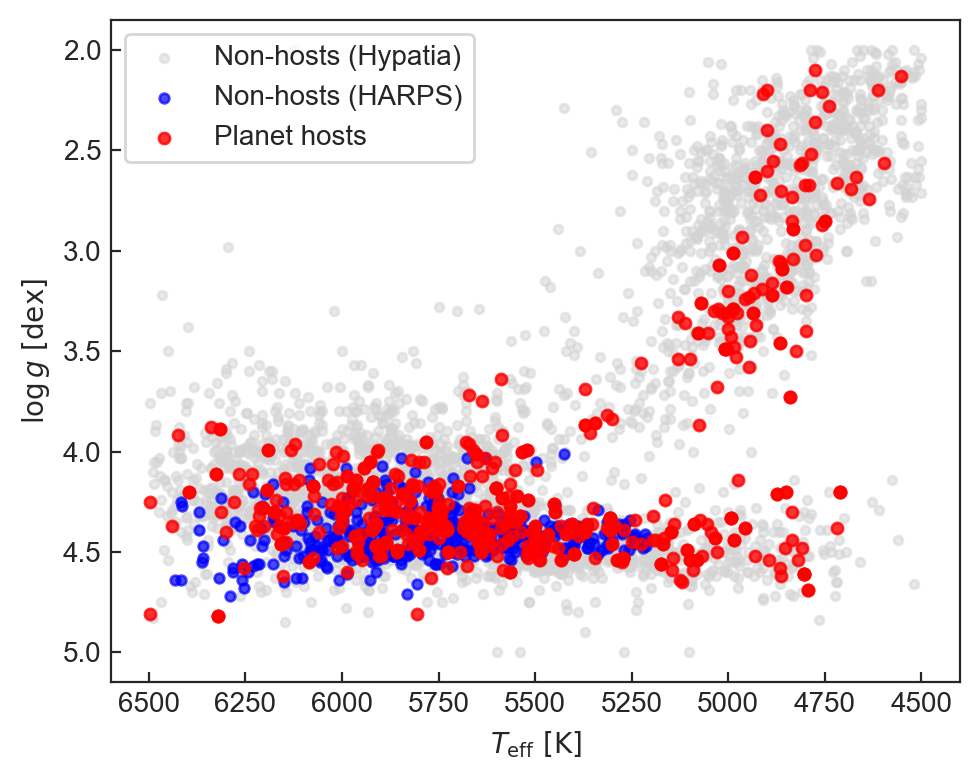}
    \caption{Distribution of stars in the Kiel diagram. Grey points represent the non-host sample from Hypatia, blue points indicate the HARPS non-host stars, and red points correspond to the planet-hosting stars from our sample.
    }
    \label{fig:kiel_diagram}
\end{figure*}

The distribution of stars with and without planets in the surface gravity–effective temperature plane (Kiel diagram) is shown in Figure~\ref{fig:kiel_diagram}. "Machine-readable tables containing the relevant parameters of the samples are available in Zenodo at \url{https://doi.org/10.5281/zenodo.17360489}.

\section{Summed mass fraction of heavy elements -- $Z$}\label{sec:Z}

For both the planet-hosting and non-host stars, we estimated $Z$ in the planet-forming material following the method described in our previous work \citep{Nguyen-24}. We adopted the stoichiometric model developed by \cite{Santos-17b}, which estimates $Z$ based on the abundances of the five key refractory and volatile elements listed above. Using the solar reference abundances from \cite{Asplund-21}, the model predicts a solar value of $Z_{\odot} = 1.25\%$ for the planet-building blocks in the Solar System. Uncertainties in $Z$ were estimated via a Monte Carlo approach that accounts for the reported uncertainties in the individual elemental abundances.

\begin{figure*}[h]
    \centering
    \includegraphics[width=0.6\textwidth]{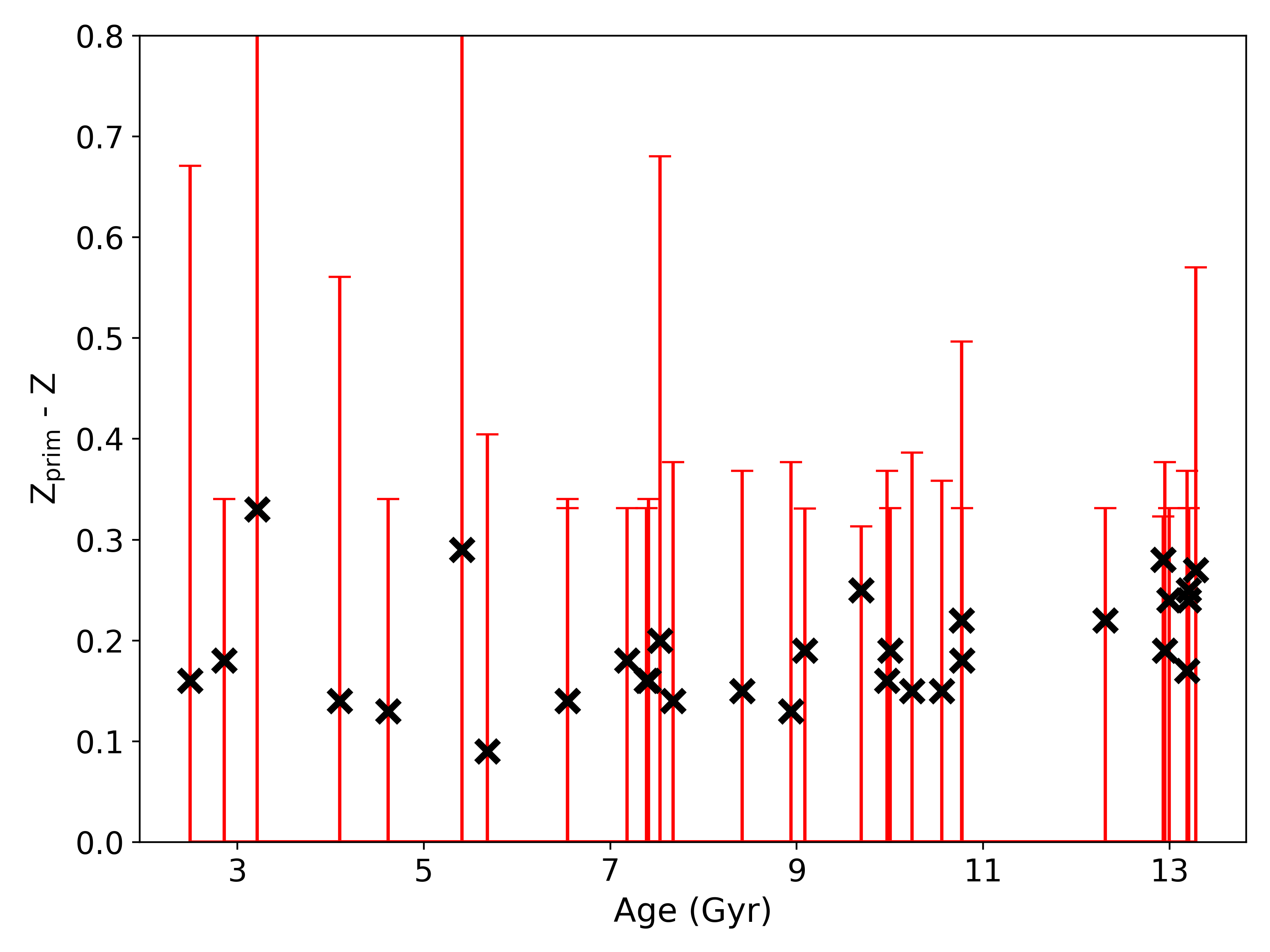}
    \caption{Difference between primordial and present-day $Z$ as a function of stellar age (crosses) for the sample of 30 FGK stars from \citet{Adibekyan-24}. The red error bars denote the uncertainties in the $Z$ differences.
    }
    \label{fig:Z_diffusion_plot}
\end{figure*}

We note that present-day stellar abundances were used as proxies for the primordial composition of protoplanetary disks. Although certain stellar processes can alter surface abundances over time \citep{Deal-18}, the relative abundances of the rock-forming elements remain largely unaffected \citep{Adibekyan-24}. Since deriving primordial compositions for every star in our sample would be computationally prohibitive, we tested the reliability of using present-day abundances by analyzing a sample of 30 FGK stars from \citet{Adibekyan-24}, for which both present-day and primordial abundances were available. From these abundances, we derived $Z$ and $Z_{\mathrm{prim}}$ and examined their difference as a function of stellar age (see Fig.~\ref{fig:Z_diffusion_plot}). We find that, on average, the difference in $Z$ corresponds to about 0.57 times the uncertainty in this parameter, and it is always smaller than the associated uncertainty. Therefore, we consider it reasonable to adopt present-day $Z$ values in our analysis. 

\subsection{Distribution of $Z$ among planet-hosting and non-host stars} \label{z_hosts_nonhosts}

\begin{figure*}[h]
    \centering
    \includegraphics[width=0.48\textwidth]{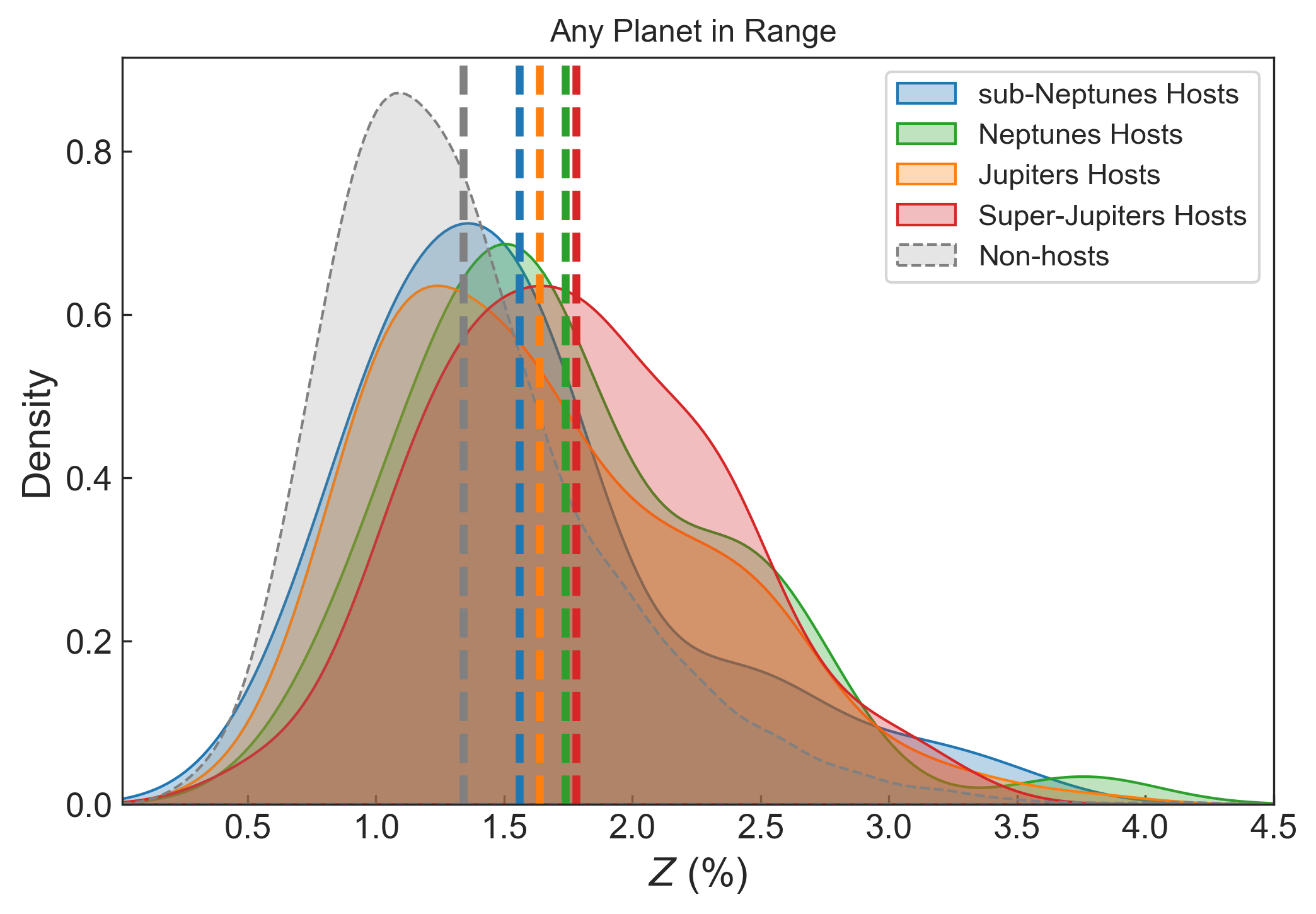}
    \hspace{0.0\textwidth}
    \includegraphics[width=0.48\textwidth]{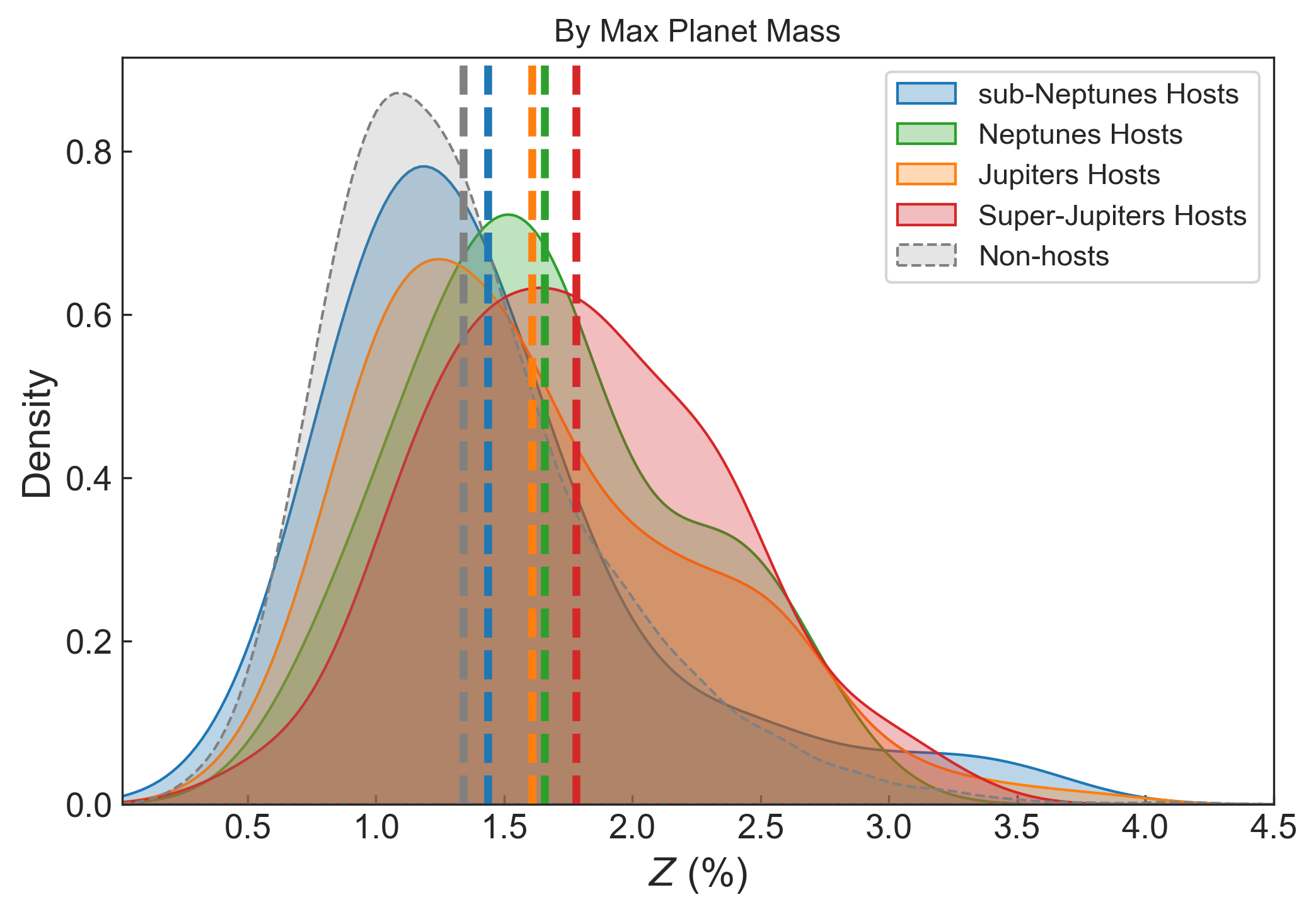}
    \caption{$Z$ distributions of planet-host and non-host stars. \textbf{Left:} Distributions based on individual planet detections, with hosts grouped by the mass of any detected planet in the system. \textbf{Right:} Same as left, but hosts are grouped according to the mass of the most massive planet in each system. The vertical dashed lines indicate the mean $Z$ of each distribution. The non-host sample is shown in gray for comparison.
    }
    \label{fig:Z_distributions}
\end{figure*}

The discovery of a correlation between the presence of giant planets and stellar iron abundance, used as a proxy for metallicity, marked a turning point in exoplanet research \citep{Gonzalez-97, Santos-04}. This finding played a pivotal role in the development of planet formation theories, particularly in supporting the core-accretion model \citep[e.g.][]{Ida-04, Mordasini-12}. In Figure~\ref{fig:Z_distributions}, we examine the distribution of $Z$ for stars with and without planets. Unlike [Fe/H], $Z$ provides a more physically meaningful measure of the solid content in protoplanetary disks relevant for planet formation.

We grouped planets by mass and considered two selection methods: one based on the presence of at least one planet within a given mass range, and another based on the most massive planet in the system falling within that range. One $Z$ value per system was used, since all planets in a system share the same stellar composition. Using  $Z$ multiple times per system would artificially inflate the weight of multiplanetary systems, which could bias the analysis, particularly if planetary multiplicity is correlated with metallicity, as discussed later in the article. To test whether the $Z$ distributions of planet-hosting and non-host stars differ, we applied Kolmogorov--Smirnov (KS) and Anderson--Darling (AD) statistical tests\footnote{We used \texttt{scipy.stats.anderson\_ksamp} to compare the distributions, where the returned p-values are capped at 0.25 and floored at 0.001.}.

Figure~\ref{fig:Z_distributions} and Table~\ref{tab:Z_combined_stats} show that, across all planet-host samples, stars hosting planets in any given mass range generally exhibit higher mean $Z$ values than non-host stars from the Hypatia Catalog. The mean $Z$ of Neptune hosts is slightly higher than that of Jupiter hosts, although this difference is not statistically significant (see Sect.~\ref{z_hosts}) and likely reflects the relatively small number of Neptune hosts, as also indicated by their larger standard error of the mean.

When comparing sub-Neptune hosts to the HARPS non-host sample, the KS and AD tests yield smaller $p$-values (KS = 0.009; AD $< 0.001$), reinforcing the significance of the difference in their metallicity distributions. In contrast, systems where the most massive planet is sub-Neptune-like show $Z$ distributions that are statistically indistinguishable from those of non-hosts, even when compared with the HARPS control sample (KS = 0.32; AD = 0.19). These results suggest that the formation of exclusively low-mass planets may not require particularly metal-rich disks, consistent with predictions from core accretion theory \citep[e.g.,][]{Mordasini-12, Hasegawa-14}.

\begin{table}[h]
\centering
\caption{Metallicity comparison between planet hosts and non-hosts. Mean and standard error of the mean of $Z$ values, along with statistical test results comparing planet-host samples to non-host stars from the Hypatia catalog. The planet-host samples were defined using two selection strategies: one based on the presence of any planet within a given mass range, and another based on the maximum planet mass in each system. The reported p-values are from the KS and AD tests.}
\label{tab:Z_combined_stats}
\begin{tabular}{lccc}
\hline
\textbf{Sample} & \textbf{Mean $Z$ (\%)} & \textbf{KS p-value} & \textbf{AD p-value} \\
\hline
sub-Neptunes (any)             & $1.56 \pm 0.09$ & 0.060 & 0.009 \\
Neptunes (any)           & $1.77 \pm 0.10$ & $2.22 \times 10^{-4}$ & $< 0.001$ \\
Jupiters (any)           & $1.64 \pm 0.04$ & $9.39 \times 10^{-10}$ & $< 0.001$ \\
Super-Jupiters (any)     & $1.78 \pm 0.06$ & $6.59 \times 10^{-12}$ & $< 0.001$ \\
\hline
sub-Neptunes (max mass)        & $1.44 \pm 0.10$ & 0.64 & $> 0.25$ \\
Neptunes (max mass)      & $1.66 \pm 0.10$ & 0.006 & $< 0.001$ \\
Jupiters (max mass)      & $1.61 \pm 0.04$ & $9.75 \times 10^{-7}$ & $< 0.001$ \\
Super-Jupiters (max mass)& $1.78 \pm 0.06$ & $6.59 \times 10^{-12}$ & $< 0.001$ \\
\hline
Non-hosts (Hypatia)               & $1.34 \pm 0.01$ & -- & -- \\
Non-hosts (Harps)               & $1.28 \pm 0.02$ & -- & -- \\
\hline
\end{tabular}
\end{table}

\subsection{Distribution of $Z$ among planet-hosting stars} \label{z_hosts}

The two control samples used in this work, namely the broader Hypatia Catalog and the more homogeneous HARPS survey, naturally differ in their selection functions and completeness. Nevertheless, the fact that the $Z$ distributions of planet hosts yield consistent results when compared with both samples increases confidence in the robustness of our conclusions. This consistency suggests that, despite the possible presence of undetected low mass or long period planets in either catalog, the overall trends remain reliable.

In addition, comparing the metallicity distributions of hosts of different planet types may provide  insights into how $Z$ relates to planetary mass. To ensure statistical validity, we restrict our analysis to the most massive planet in each system because the KS and AD tests assume independence of the two samples, and including the same star in multiple groups would violate this assumption.

When comparing the host populations under this criterion, we find that sub-Neptune hosts are less metal rich than Neptune hosts ($p < 0.04$) and super-Jupiter hosts ($p < 0.001$), but they are statistically indistinguishable from Jupiter hosts ($p \sim 0.09$) if considering the 0.05 threshold for p-values. Neptune hosts exhibit $Z$ distributions similar to those of both Jupiter hosts ($p > 0.25$) and super-Jupiter hosts ($p > 0.19$), while the $Z$ distribution of Jupiter hosts is significantly lower than that of super Jupiter hosts ($p < 0.01$). At first sight these results may seem counterintuitive since sub-Neptunes are less metal rich than Neptunes yet statistically similar to Jupiters, while Neptunes and Jupiters also appear similar. This pattern arises from overlapping distributions and modest effect sizes, with sample variance causing some pairwise tests to fall just above or just below the significance threshold. Overall, the mean metallicities support a general increase from sub-Neptunes to super-Jupiters.

\subsection{Planetary system multiplicity as a function of $Z$} \label{multi_single}

\begin{figure*}[h]
    \centering
    \includegraphics[width=0.32\textwidth]{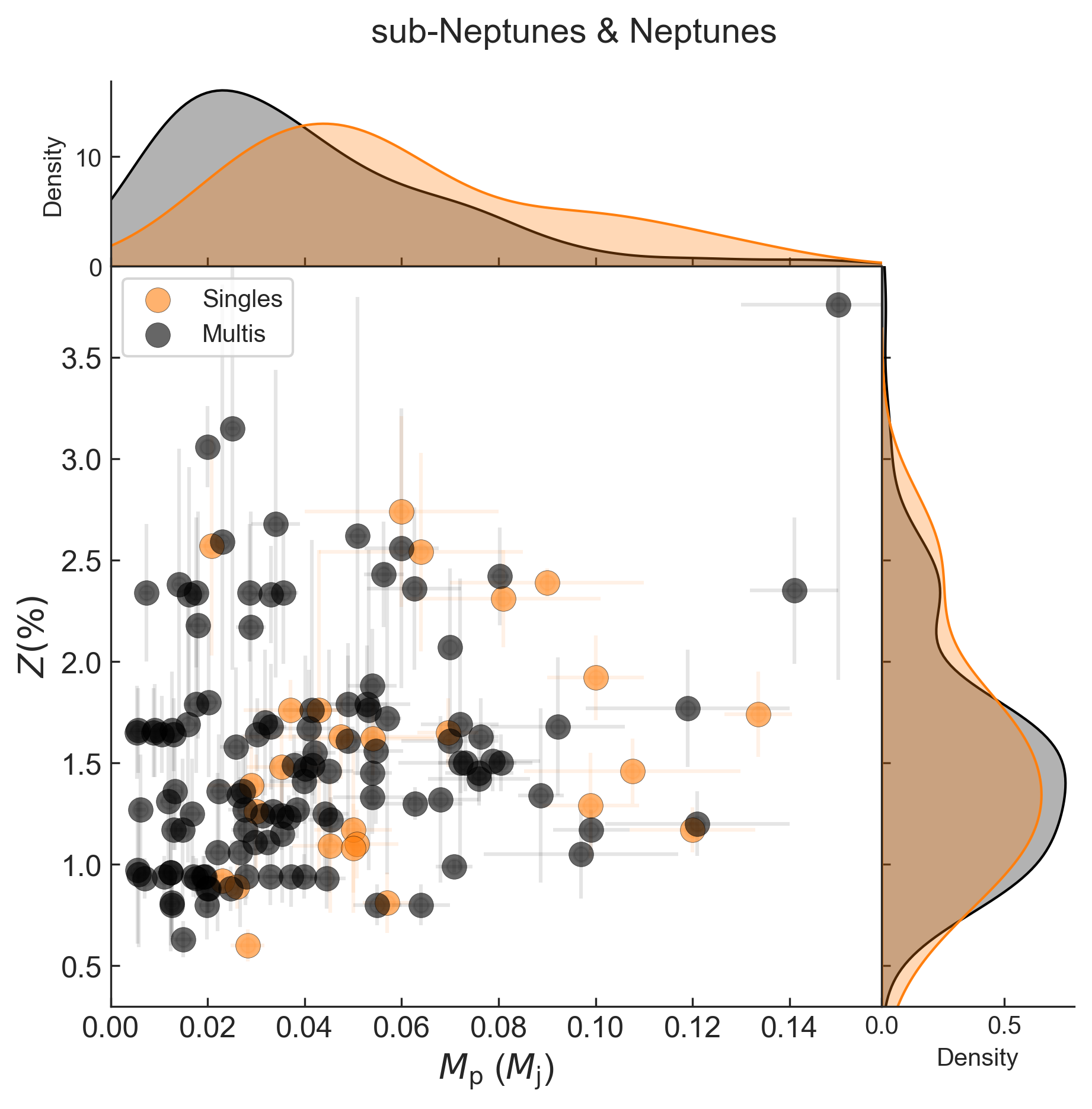}
    \hspace{0.0\textwidth}
    \includegraphics[width=0.32\textwidth]{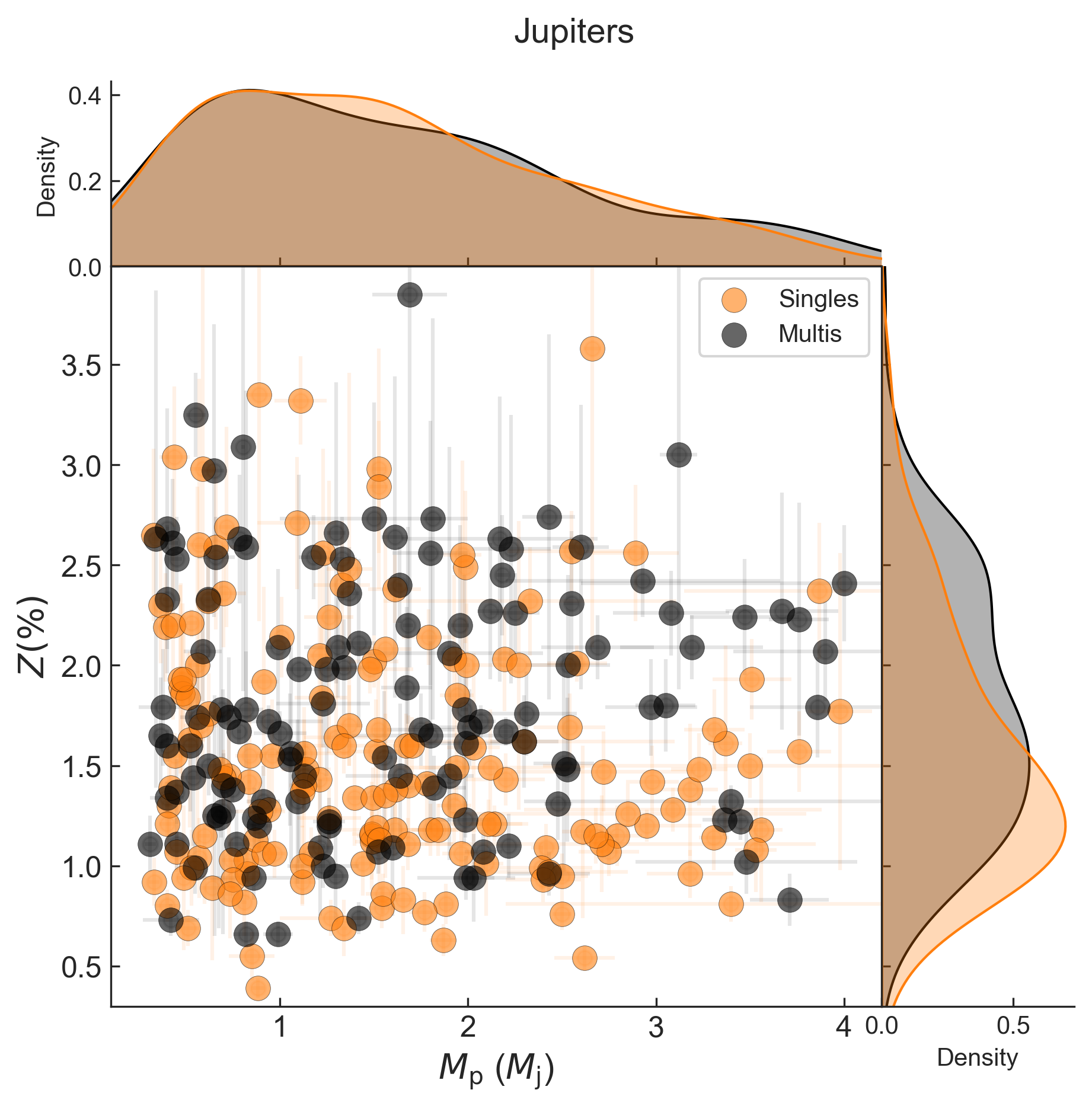}
    \hspace{0.0\textwidth}
    \includegraphics[width=0.32\textwidth]{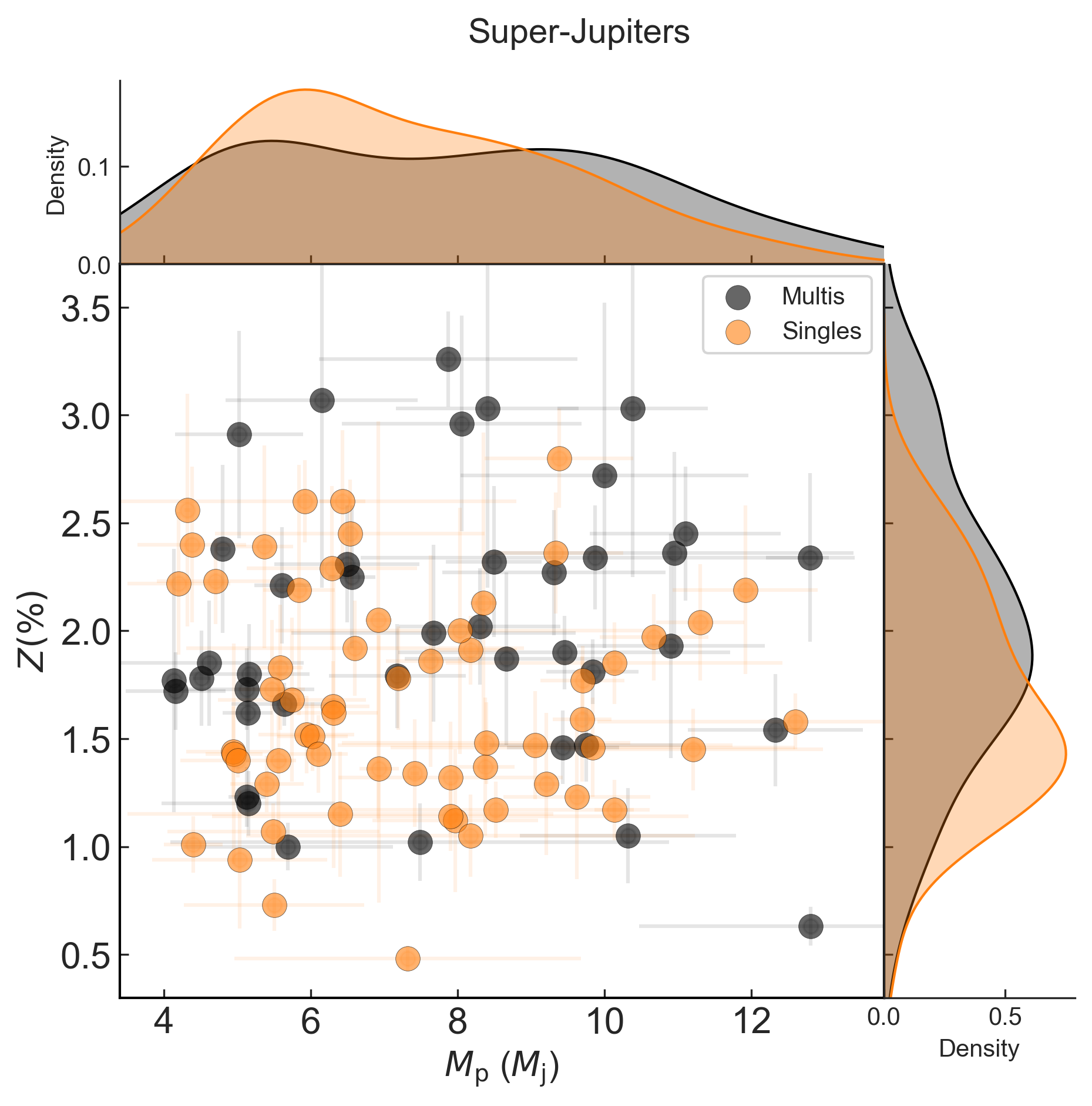}
    \caption{Relationship between $Z$ and planetary mass across system architectures. \textbf{Left:} sub-Neptune- and Neptune-mass planets. \textbf{Center:} Jupiter-mass planets. \textbf{Right:} Super-Jupiter-mass planets. Black points correspond to multiplanetary systems, while orange points indicate single-planet systems. Smoothed marginal distributions via Kernel density estimates (KDEs) of $Z$ and $M_{\mathrm{p}}$ are shown for both populations.
    }
    \label{fig:Z_vs_mass_multiplicity}
\end{figure*}

Since the early studies \citep{Fischer-05}, it has been suggested that metal-rich stars are more likely to host multiple giant planets \citep[e.g.][]{Buchhave-18}, a trend naturally explained by the CA model \citep[e.g.][]{Wimarsson-20}. However, these studies typically relied on [Fe/H] as a proxy for overall metallicity and focused primarily on Jupiter-mass planets orbiting Sun-like stars \citep{Buchhave-18} or on small samples of evolved stars \citep{Teng-23}. In Figure~\ref{fig:Z_vs_mass_multiplicity}, we extend this analysis by investigating the impact of $Z$ on the multiplicity of planetary systems across different planet mass regimes, including the super-Jupiters. The comparative properties of single- and multi-planet systems, together with the statistical test results on the $Z$ distributions of these samples, are summarized in Table~\ref{tab:multiplicity_stats_improved}.

\begin{table}[h]
\centering
\caption{Differences between single and multiplanetary systems. Comparison of system properties between single and multiplanetary systems across different planetary mass regimes. Mean values and standard errors of the means are shown, along with p-values from the KS and AD tests.}
\label{tab:multiplicity_stats_improved}
\begin{tabular}{llcccc}
\hline
\textbf{Mass Regime} & \textbf{Property} & \textbf{Single} & \textbf{Multi} & \textbf{KS p-value} & \textbf{AD p-value} \\
\hline
\multirow{4}{*}{sub-Neptune+Neptune} 
 & $Z$ (\%)                & $1.55 \pm 0.06$     & $1.63 \pm 0.11$     & 0.98   & $> 0.25$ \\
 & Period (days)          & $74 \pm 30$    & $77 \pm 17$    & 0.21   & $> 0.25$ \\
 & Eccentricity           & $0.21 \pm 0.04$     & $0.15 \pm 0.02$     & 0.11   & 0.15 \\
 & Mass ($M_{\mathrm{J}}$)& $0.060 \pm 0.006$     & $0.040 \pm 0.003$     & 0.011  & $< 0.001$ \\
\hline
\multirow{4}{*}{Jupiter}       
 & $Z$ (\%)                & $1.55 \pm 0.05$     & $1.82 \pm 0.07$     & 0.003  & $< 0.001$ \\
 & Period (days)          & $1135 \pm 204$ & $1596 \pm 227$ & 0.32   & 0.08 \\
 & Eccentricity           & $0.21 \pm 0.02$     & $0.21 \pm 0.02$     & 0.97   & $> 0.25$ \\
 & Mass ($M_{\mathrm{J}}$)& $1.58 \pm 0.07$     & $1.60 \pm 0.09$     & 0.99   & $> 0.25$ \\
\hline
\multirow{4}{*}{Super-Jupiter} 
 & $Z$ (\%)                & $1.67 \pm 0.07$     & $1.96 \pm 0.10$     & 0.03  & 0.02 \\
 & Period (days)          & $3760 \pm 858$ & $7164 \pm 1389$ & $< 0.001$  & $< 0.001$ \\
 & Eccentricity           & $0.35 \pm 0.03$     & $0.24 \pm 0.03$     & 0.04  & 0.03 \\
 & Mass ($M_{\mathrm{J}}$)& $7.33 \pm 0.27$     & $7.82 \pm 0.40$     & 0.41   &  $> 0.25$ \\
\hline
\end{tabular}
\end{table}

In the left panel of Figure~\ref{fig:Z_vs_mass_multiplicity}, we present the combined sample of sub-Neptune- and Neptune-mass planets, comprising 26 single-planet systems and 57 multiplanetary systems. These two categories were merged due to the small number of single-planet systems, particularly in the case of sub-Neptunes.

While the average $Z$ of multiplanetary systems is slightly higher than that of single-planet systems, the difference is not statistically significant. This is supported by the p-values of the KS and AD tests, both of which exceed the commonly adopted significance threshold of 0.05. Similarly, the distributions of orbital periods and eccentricities do not differ significantly between the two populations. Our results are consistent with those obtained from the transiting planets of the California-Kepler Survey \citep{Weiss-18, Ghezzi-21}. The only statistically significant difference identified in this mass regime is in the distribution of planetary masses: systems with multiple planets tend to host lower-mass planets on average.

Our sample includes 165 single and 80 multiple systems hosting Jupiter-mass planets. In contrast to the low-mass regime, Jupiter-mass planets in multiplanetary systems are found in systematically more metal-rich environments than those in single-planet systems, with p-values of $< 0.003$ from both the KS and AD tests. The distributions of orbital periods, eccentricities, and planetary masses show no statistically significant differences.

Finally, our sample includes 60 single and 36 multiple systems hosting super-Jupiter-mass planets. As in the case of Jupiters, super-Jupiters in multiplanetary systems are found in systematically more metal-rich environments than those in single-planet systems, with $p$-values of $\sim 0.03$ from both the KS and AD tests. Notably, the mean $Z$ values for both single and multiple systems in the super-Jupiter regime are higher than those of Jupiter-mass planets, although the difference is statistically significant ($p < 0.03$) only for the single systems. These results reinforce the view that heavy-element enrichment remains important even at the high end of the planet-mass spectrum.

While the distributions of planetary masses are statistically similar, we observe  differences in orbital properties. Super-Jupiters in multiplanetary systems tend to have longer orbital periods (p-values $<$ 0.001) and lower eccentricities (p-values $<$ 0.04). These values are also larger than those observed for Jupiter-mass planets, both in period and eccentricity (p-values $<$ 0.01).

The elevated eccentricities of the most massive single planets have been noted in earlier studies \citep[e.g.][]{Adibekyan-13} and are interpreted as a consequence of planet--disk interactions during formation and migration \citep{Bitsch-13}. The lower eccentricities observed in multiplanetary systems may reflect the presence of lower-mass companions; since eccentricity tends to decrease with planet mass, the average eccentricity across such systems would naturally be lower.

\subsection{Relationship between $Z$ and planetary masses}

To further investigate how the heavy-element content of protoplanetary disks influences planet formation outcomes, we examine correlations between $Z$ and planetary masses. Early studies suggested that the total planetary mass in high-mass systems may increase with stellar metallicity \citep{Fischer-05}. Subsequent works indicated that the average host star metallicity, as traced by [Fe/H], tends to rise with increasing planet mass—although this trend appears to reverse for super-Jupiters \citep{Adibekyan-13, Narang-18, Maldonado-19}. More recently, considering total metal content in the disks, a weak positive correlation was reported between planetary mass and disk metallicity for planets in the range 1–13\,$M_{\mathrm{J}}$ \citep{Nguyen-24}.

\begin{figure*}[h]
    \centering
    \includegraphics[width=0.32\textwidth]{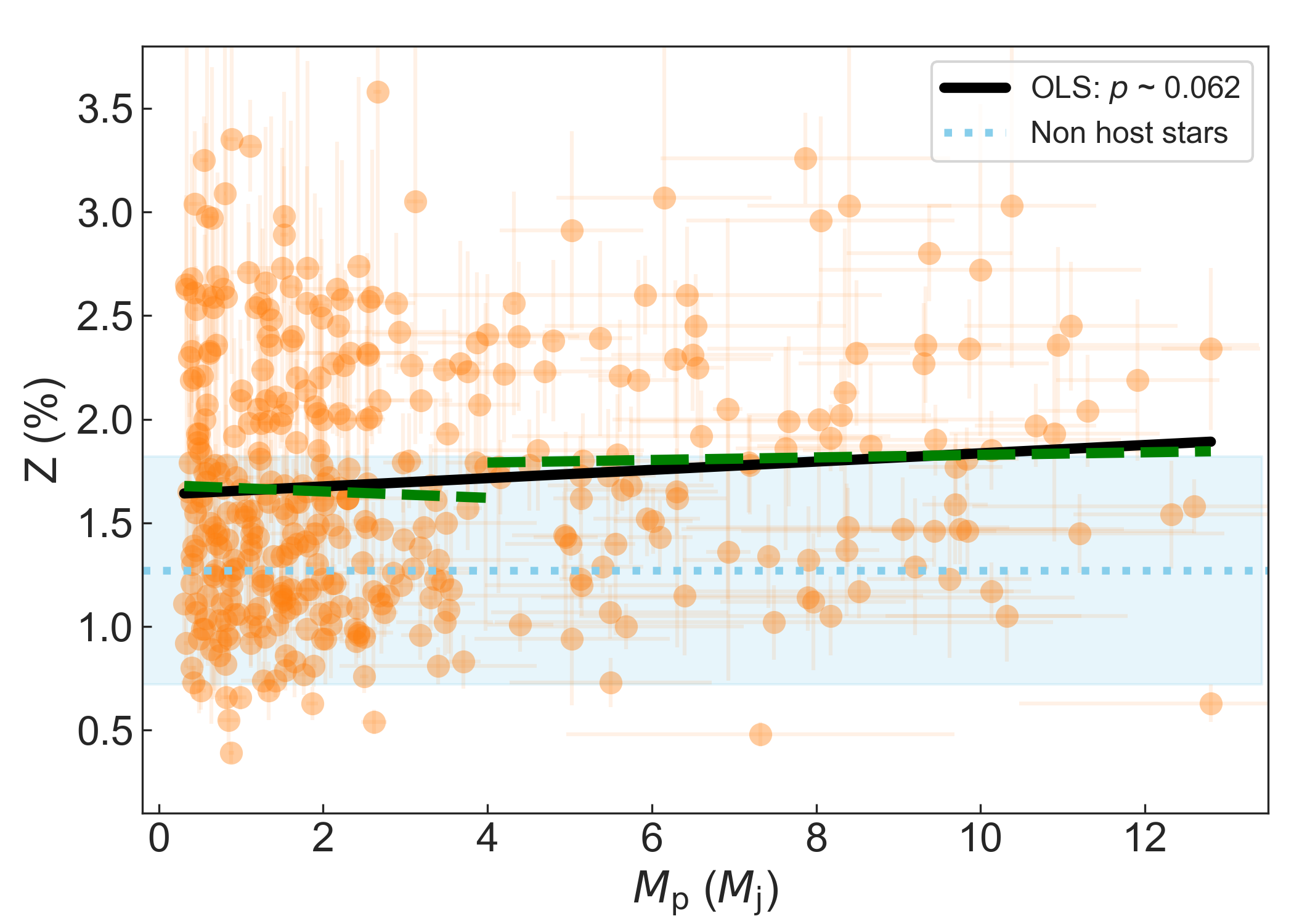}
    \hspace{0.0\textwidth}
    \includegraphics[width=0.32\textwidth]{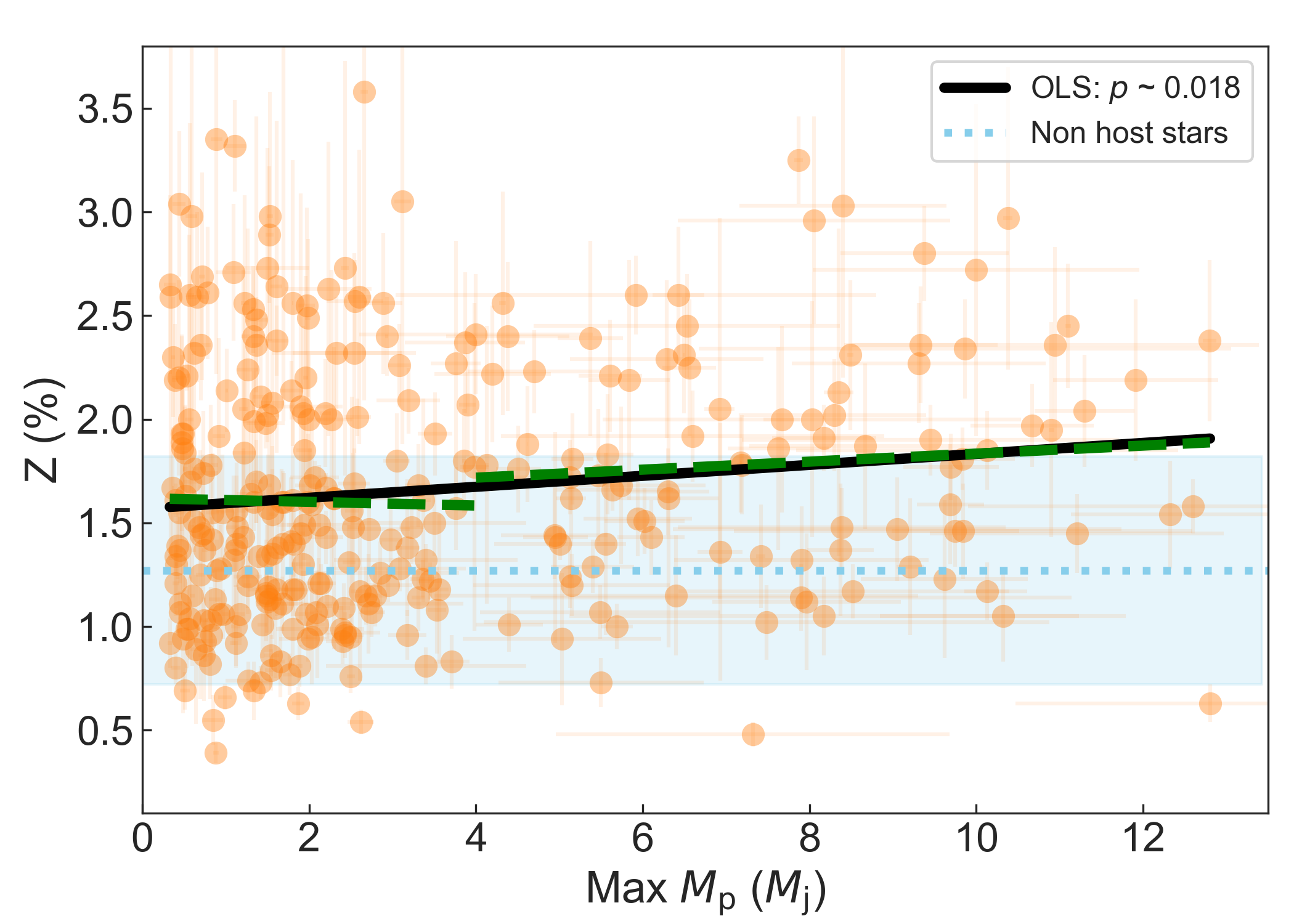}
    \hspace{0.0\textwidth}
    \includegraphics[width=0.32\textwidth]{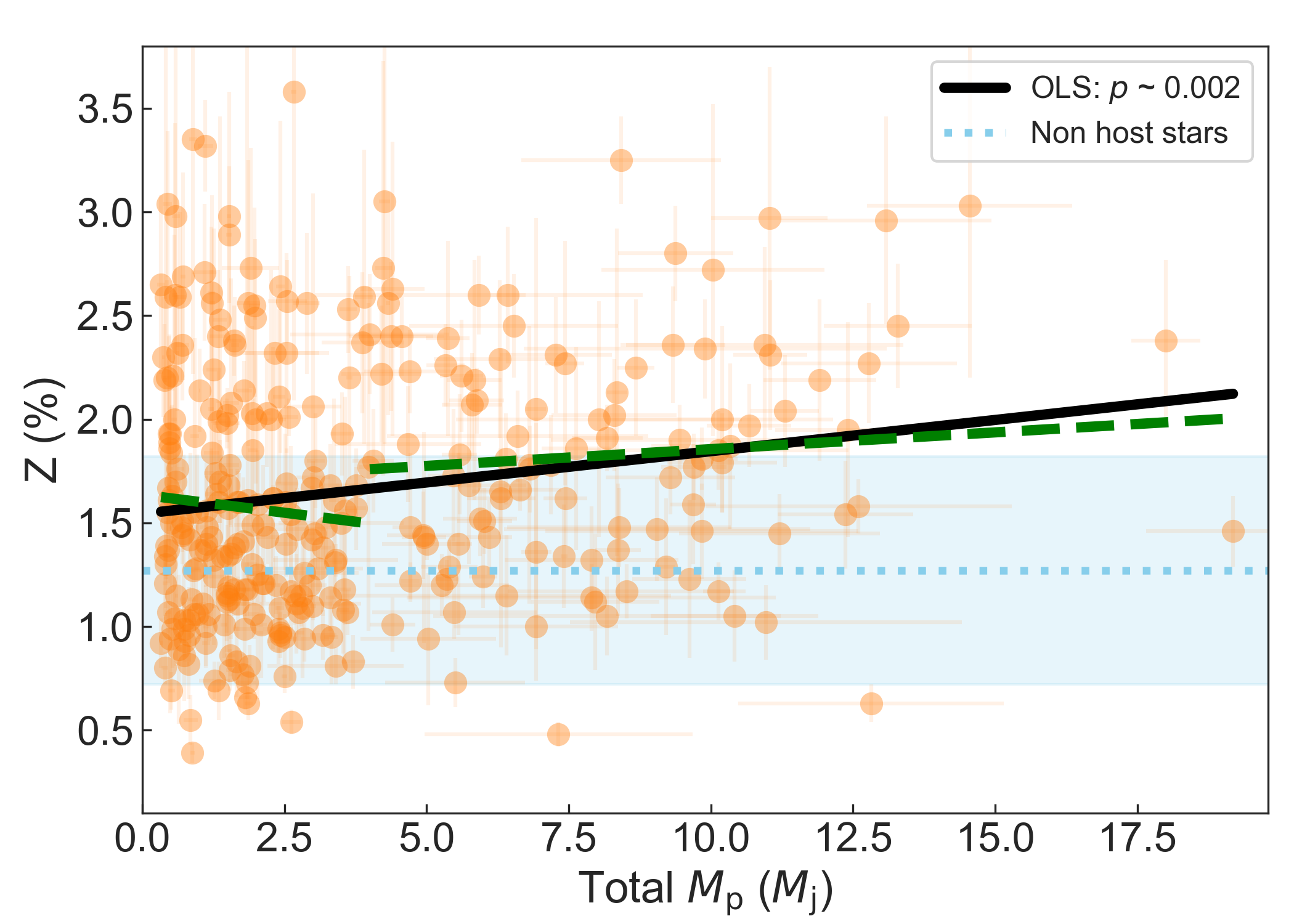}
    \caption{Summed mass fraction of heavy elements as a function of planetary mass.
     \textbf{Left:} $Z$ versus the mass of individual planets in the system. \textbf{Center:} $Z$ versus the mass of the most massive planet in each system. \textbf{Right:} $Z$ versus the total planetary mass in the system.  The solid black line shows the best-fit OLS regression to the full sample. Green dashed lines show separate OLS fits to the Jupiter-mass and super-Jupiter-mass regimes. The horizontal dotted blue line indicates the mean $Z$ of stars without detected planets, with the shaded region showing the $\pm1\sigma$ interval. 
    }
    \label{fig:Z_vs_mass}
\end{figure*}

In the left panel of Figure~\ref{fig:Z_vs_mass}, we present the relation between $Z$ and planetary mass for Jupiter- and super-Jupiter-mass planets. The slope of the best-fit linear relation, obtained via ordinary least squares (OLS) regression, is small ($0.020 \pm 0.011$), and the $p$-value ($\sim 0.06$) lies slightly above the commonly adopted significance threshold of 0.05. However, this result may be influenced by the correlation between metallicity and planetary system multiplicity. In particular, if the likelihood of hosting multiple planets depends on both stellar metallicity and planet mass, then the presence of several planets from the same system in the $Z$–mass plane could bias the regression analysis.

To mitigate this potential bias, we also examine $Z$ as a function of the mass of the most massive planet in the system (middle panel) and the total planetary mass per system (right panel). The fitted slopes remain shallow: $0.026 \pm 0.011$ for maximum planet mass and $0.030 \pm 0.010$ for total planetary mass, but the statistical significance increases, with p-values of 0.02 and 0.002, respectively. 

A closer inspection of the plots reveals that these trends may be primarily driven by the most massive planets. The green dashed lines, representing separate OLS fits for the Jupiter- and super-Jupiter-mass regimes, indicate that the slope for Jupiters is nearly flat or even negative, while a weak positive correlation is seen for super-Jupiters. For Jupiter-mass planets, the OLS slopes are $-0.015 \pm 0.040$ ($p = 0.70$) for $Z$–$M_{\mathrm{p}}$, $-0.000 \pm 0.045$ ($p = 0.85$) for $Z$–Max $M_{\mathrm{p}}$, and $-0.034 \pm 0.045$ ($p = 0.43$) for $Z$–Total $M_{\mathrm{p}}$. For super-Jupiters, the corresponding slopes are $0.010 \pm 0.025$ ($p = 0.68$), $0.025 \pm 0.025$ ($p = 0.34$), and $0.019 \pm 0.019$ ($p = 0.31$). None of these subsample trends is statistically significant, as all $p$-values lie well above the 0.05 threshold. However, the correlations become statistically significant when the lower mass limit of the super-Jupiter regime is reduced to $\sim$2 $M_{\mathrm{J}}$, suggesting that the apparent positive trend is mainly driven by planets in the $>$ 2 $M_{\mathrm{J}}$ regime.

To assess the reliability of our findings, we next examine the robustness of our results under different uncertainty thresholds.

\section{Robustness checks}

Throughout this study we adopted several working assumptions and performed different statistical tests. In particular, when selecting the samples we required relative uncertainties in both $Z$ and planetary masses to be below 50\%. Here we also tested alternative thresholds to verify that our results are not sensitive to the exact choice of this limit. In the following, we discuss the impact of these assumptions on the main results.

As an alternative to the 50\% threshold, we applied more stringent cuts on the relative uncertainties. Reducing the mass uncertainty cut to 15\% decreased the sample to 428 planets around 321 stars, while still maintaining at least 20 hosts in each planetary mass group. Similarly, restricting the relative uncertainty in $Z$ to below 20\% yielded a sample of 291 hosts with 393 planets (again with at least 20 hosts in each group), 3,231 non-host stars from the Hypatia Catalog, and 233 stars from the HARPS sample. We note that the variations in the precision cuts for planetary mass and $Z$ were performed independently.

In both cases, limiting the analysis to the highest-precision subsets did not alter the main results concerning the comparison of $Z$ distributions between stars with and without planets, although the exact $p$-values changed.

We also repeated the tests related to multiplicity and $Z$ using the restricted samples. While the exact $p$-values again shifted, the overall conclusions remained unchanged for low-mass and Jupiter-mass planets. For super-Jupiters, however, the sample size was substantially reduced. Although the difference in mean $Z$ values between single and multiple systems persisted (mass cut: $1.97 \pm 0.12$ vs. $1.74 \pm 0.08$; $Z$ cut: $1.89 \pm 0.11$ vs. $1.70 \pm 0.08$), the corresponding distributions were no longer statistically distinct ($p > 0.15$).

Finally, we tested the $Z$–planetary mass relations using the restricted samples. Despite some variation in the $p$-values, all of the observed trends remained unchanged.

Overall, these tests demonstrate that our conclusions are robust against the adopted uncertainty thresholds in both planetary masses and $Z$.

\section{[Fe/H] as a proxy for metallicity}

Throughout this paper we used $Z$ as a more astrophysically motivated proxy of metallicity for planet formation than iron abundance alone. In this section, we repeat the main analysis using [Fe/H] instead of $Z$. The mean [Fe/H] values of the samples are listed in Table~\ref{tab:FeH_combined_stats}, showing close similarity to the corresponding $Z$ values. The main difference is that non-hosts from the HARPS sample appear significantly more metal-poor than their counterparts in the Hypatia catalog. However, this offset does not alter the overall conclusions obtained with $Z$. In particular, the comparison of [Fe/H] distributions between planet hosts and non-hosts yields results consistent with those discussed in Sect.~\ref{z_hosts_nonhosts}, with differences limited to the exact $p$-values.

\begin{table}[h]
\centering
\caption{Mean [Fe/H] values of planet-host samples and non-host stars. The planet-host samples are defined using two selection strategies: one based on the presence of any planet within a given mass range, and another based on the maximum planet mass in each system. Reported values are the mean and standard error of the mean.}
\label{tab:FeH_combined_stats}
\begin{tabular}{lcc}
\hline
\textbf{Sample} & \textbf{[Fe/H] (any)} & \textbf{[Fe/H] (max)} \\
\hline
sub-Neptunes       & $-0.00 \pm 0.03$ & $-0.06 \pm 0.04$ \\
Neptunes           & $0.12 \pm 0.03$  & $0.12 \pm 0.03$ \\
Jupiters           & $0.09 \pm 0.01$  & $0.08 \pm 0.01$ \\
Super-Jupiters     & $0.14 \pm 0.02$  & $0.14 \pm 0.02$ \\
\hline
Non-hosts (Hypatia) & $-0.03 \pm 0.005$ & -- \\
Non-hosts (HARPS)   & $-0.12 \pm 0.01$  & -- \\
\hline
\end{tabular}
\end{table}

The results regarding multiplicity and the relation between metallicity and planetary mass also remain broadly unchanged, despite shifts in the exact $p$-values. The only notable difference is that when using [Fe/H], the slopes of the [Fe/H]–planetary mass relation for low-mass Jupiters ($0.3 < M_{\mathrm{p}} \leq 4 M_{\mathrm{J}}$) are never negative, although they remain statistically insignificant.

It is important to emphasize that the similarity of results obtained with [Fe/H] and $Z$, while reassuring, should not be interpreted as validation of iron abundance alone as a reliable tracer of disk metallicity in the context of planet formation. For example, the Hypatia and HARPS non-host samples show statistically indistinguishable $Z$ distributions (KS $p \sim 0.15$), yet their [Fe/H] distributions differ at high significance (KS $p \sim 6 \times 10^{-9}$). This highlights the limitations of relying solely on [Fe/H]. Ideally, future studies should employ larger samples of planet hosts with homogeneously determined abundances of the key refractory and volatile elements to better constrain the connection between stellar composition and planet formation.

\section{Discussion and conclusion}

Since the discovery of the first exoplanet orbiting a Sun-like star \citep{Mayor-95}, our understanding of planet formation has evolved beyond the framework established by studies of the Solar System. Two main theoretical mechanisms have been proposed to explain the formation of planets: the CA model and the GI,  both of which have experienced substantial development and modifications during the last decades. These mechanisms may operate in competition or complementarity, depending on the local conditions in the protoplanetary disk \citep{Boss-97, Marois-10, Mordasini-12, Savvidou-23}. It has been also proposed that  if the GI is followed by tidal stripping during the migration of the clumps, then even a rocky planet could form \citep{Nayakshin-17}

There is general consensus that CA is the dominant mechanism responsible for forming both low-mass and Jupiter-mass planets. In contrast, the formation of the most massive planets  remains a subject of debate. It has been proposed that such massive planets may form via GI, due to the seemingly weaker or opposite metallicity dependence observed in earlier studies. This interpretation has led to the view that CA and GI operate in distinct domains, with GI becoming dominant at the upper end of the planetary mass spectrum.

By using an astrophysically motivated summed mass fraction of heavy elements $Z$, we find several lines of evidence suggesting that even the most massive planets can form via core accretion. First, the $Z$ distribution of super-Jupiter hosts is statistically distinct from that of non-host stars and shows even higher average value than the Jupiter host population, a key prediction of CA models \citep{Mordasini-12}. Second, multiplanetary systems containing super-Jupiters are more metal-rich (high-$Z$) than their single-planet counterparts. However, the significance of this results drops when considering a restricted sample with the highest precision in planetary mass and $Z$. The multiplanetary systems also exhibit significantly higher total planetary masses (multis: $14.2 \pm 1.5\,M_{\mathrm{J}}$ vs singles: $7.3 \pm 0.3\,M_{\mathrm{J}}$), with p-values from both KS and AD tests below 0.001. A similar trend is observed among Jupiter systems (multis: $7.1 \pm 0.4\,M_{\mathrm{J}}$ vs singles: $1.6 \pm 0.07\,M_{\mathrm{J}}$). These findings reinforce the interpretation that metal enrichment facilitates the formation of massive planets, consistent with the expectations of CA. Finally, we observe a tentative positive correlation between $Z$ and both the total and maximum planetary mass in a system. This trend appears to be driven primarily by the most massive systems and aligns with theoretical predictions that, while metallicity acts as a threshold for giant planet formation, it does not strongly correlate with mass except at the high end of the planet mass spectrum \citep{Mordasini-12}.

While our findings highlight the central role of heavy-element content in shaping planetary architectures, they do not rule out the possibility that GI may operate under specific conditions. This mechanism may be especially relevant for the formation of super-Jupiters around M dwarfs, where CA becomes increasingly inefficient and the formation of such massive planets may instead require disk fragmentation through GI \citep{Boss-24}. A few super-Jupiter-mass planets are found in subsolar $Z$ environments (see Figure~\ref{fig:Z_vs_mass}), which could point toward GI. However, these are RV-detected planets, and their true masses may exceed the adopted $13\,M_{\mathrm{J}}$ threshold for classifying planets, thus potentially falling outside the planetary regime altogether.

Overall, our results reveal a coherent trend between host-star $Z$ and planetary properties across the full spectrum of planetary masses, from Earth-mass planets to the most massive super-Jupiters. This observed continuity, combined with the well-established link between metallicity and CA efficiency, suggests that CA is the dominant formation mechanism shaping most observed planetary architectures. In the absence of a clear metallicity signature associated with GI-formed planets, its role appears limited to rare or extreme cases, such as the formation of massive planets at wide separations around low-mass stars or the formation of objects at the edge of the brown dwarf regime. Future observations and theoretical modeling will be crucial to disentangling the relative contributions of these two formation pathways.

While our analysis shows that the main results are qualitatively reproduced when using [Fe/H], we argue that $Z$ is the more appropriate parameter to investigate planet formation. Unlike iron abundance alone, $Z$ reflects the total heavy-element content available in the protoplanetary disk. The consistency between the two proxies strengthens the robustness of our conclusions, but the physical interpretation clearly favors $Z$ as the more appropriate measure of disk metallicity.


\begin{acknowledgments}
This research has made use of the NASA Exoplanet Archive, which is operated by the California Institute of Technology, under contract with the National Aeronautics and Space Administration under the Exoplanet Exploration Program.
The research shown here acknowledges use of the Hypatia Catalog Database, an online compilation of stellar abundance data as described in \cite{Hinkel-14}, which was supported by NASA's Nexus for Exoplanet System Science (NExSS) research coordination network and the Vanderbilt Initiative in Data-Intensive Astrophysics (VIDA).
V.~A. acknowledges support from FCT – Funda\c{c}\~ao para a Ci\^encia e Tecnologia through national funds and from FEDER via COMPETE2020 – Programa Operacional Competitividade e Internacionalização, under the grants UIDB/04434/2020 (DOI: 10.54499/UIDB/04434/2020) and UIDP/04434/2020 (DOI: 10.54499/UIDP/04434/2020), as well as through a work contract funded by the FCT Scientific Employment Stimulus program (reference 2023.06055.CEECIND/CP2839/CT0005, DOI: 10.54499/2023.06055.CEECIND/CP2839/CT0005)
\end{acknowledgments}

\begin{contribution}

M.N. conducted most of the research/calculations and drafted the manuscript. V.A. provided guidance at key stages of the project and revisions of the manuscript. Both authors approved the final version for submission.

\end{contribution}

%

\software{In this work we used the Python language and several scientific packages: \textsc{SciPy} \cite{Virtanen-20}, \textsc{Matplotlib} \cite{Hunter-07}, \textsc{Statsmodels} \cite{Seabold-10}, \textsc{Pandas} \cite{mckinney2010data}, \textsc{Seaborn} \cite{Waskom2021}. To determine the summed mass fraction of heavy elements, we used the publicly available code  at 
\href{https://github.com/vadibekyan/planet_building_blocks}{https://github.com/vadibekyan/planet\_building\_blocks}
          } \citep{planet_building_blocks}.



\bibliography{sample701}{}

\begin{thebibliography}{}
\expandafter\ifx\csname natexlab\endcsname\relax\def\natexlab#1{#1}\fi
\providecommand{\url}[1]{\href{#1}{#1}}
\providecommand{\dodoi}[1]{doi:~\href{http://doi.org/#1}{\nolinkurl{#1}}}
\providecommand{\doeprint}[1]{\href{http://ascl.net/#1}{\nolinkurl{http://ascl.net/#1}}}
\providecommand{\doarXiv}[1]{\href{https://arxiv.org/abs/#1}{\nolinkurl{https://arxiv.org/abs/#1}}}

\bibitem[{V. Adibekyan(2025)Adibekyan}]{planet_building_blocks}
Adibekyan, V. 2025, planet\_building\_blocks,,
  \url{https://github.com/vadibekyan/planet_building_blocks}

\bibitem[{V. {Adibekyan} {et~al.}(2018){Adibekyan}, {Sousa}, \&
  {Santos}}]{Adibekyan-18}
{Adibekyan}, V., {Sousa}, S.~G., \& {Santos}, N.~C. 2018,
  \bibinfo{title}{{Characterization of Exoplanet-Host Stars},} in Astrophysics
  and Space Science Proceedings, Vol.~49, Asteroseismology and Exoplanets:
  Listening to the Stars and Searching for New Worlds, ed. T.~L. {Campante},
  N.~C. {Santos}, \& M.~J.~P.~F.~G. {Monteiro}, 225,
  \dodoi{10.1007/978-3-319-59315-9_12}

\bibitem[{V. {Adibekyan} {et~al.}(2021){Adibekyan}, {Dorn}, {Sousa}, {Santos},
  {Bitsch}, {Israelian}, {Mordasini}, {Barros}, {Delgado Mena}, {Demangeon},
  {Faria}, {Figueira}, {Hakobyan}, {Oshagh}, {Soares}, {Kunitomo}, {Takeda},
  {Jofr{\'e}}, {Petrucci}, \& {Martioli}}]{Adibekyan-21}
{Adibekyan}, V., {Dorn}, C., {Sousa}, S.~G., {et~al.} 2021, \bibinfo{title}{{A
  compositional link between rocky exoplanets and their host stars},} Science,
  374, 330, \dodoi{10.1126/science.abg8794}

\bibitem[{V. {Adibekyan} {et~al.}(2024){Adibekyan}, {Deal}, {Dorn}, {Dittrich},
  {Soares}, {Sousa}, {Santos}, {Bitsch}, {Mordasini}, {Barros}, {Bossini},
  {Campante}, {Delgado Mena}, {Demangeon}, {Figueira}, {Moedas}, {Martirosyan},
  {Israelian}, \& {Hakobyan}}]{Adibekyan-24}
{Adibekyan}, V., {Deal}, M., {Dorn}, C., {et~al.} 2024,
  \bibinfo{title}{{Linking the primordial composition of planet building disks
  to the present-day composition of rocky exoplanets},} Astronomy \&
  Astrophysics, 692, A67, \dodoi{10.1051/0004-6361/202452193}

\bibitem[{V.~Z. {Adibekyan} {et~al.}(2012){Adibekyan}, {Santos}, {Sousa},
  {Israelian}, {Delgado Mena}, {Gonz{\'a}lez Hern{\'a}ndez}, {Mayor}, {Lovis},
  \& {Udry}}]{Adibekyan-12}
{Adibekyan}, V.~Z., {Santos}, N.~C., {Sousa}, S.~G., {et~al.} 2012,
  \bibinfo{title}{{Overabundance of {\ensuremath{\alpha}}-elements in
  exoplanet-hosting stars},} Astronomy \& Astrophysics, 543, A89,
  \dodoi{10.1051/0004-6361/201219564}

\bibitem[{V.~Z. {Adibekyan} {et~al.}(2013){Adibekyan}, {Figueira}, {Santos},
  {Mortier}, {Mordasini}, {Delgado Mena}, {Sousa}, {Correia}, {Israelian}, \&
  {Oshagh}}]{Adibekyan-13}
{Adibekyan}, V.~Z., {Figueira}, P., {Santos}, N.~C., {et~al.} 2013,
  \bibinfo{title}{{Orbital and physical properties of planets and their hosts:
  new insights on planet formation and evolution},} Astronomy \& Astrophysics,
  560, A51, \dodoi{10.1051/0004-6361/201322551}

\bibitem[{M. {Asplund} {et~al.}(2021){Asplund}, {Amarsi}, \&
  {Grevesse}}]{Asplund-21}
{Asplund}, M., {Amarsi}, A.~M., \& {Grevesse}, N. 2021, \bibinfo{title}{{The
  chemical make-up of the Sun: A 2020 vision},} Astronomy \& Astrophysics, 653,
  A141, \dodoi{10.1051/0004-6361/202140445}

\bibitem[{B. {Banerjee} {et~al.}(2024){Banerjee}, {Narang}, {Manoj}, {Henning},
  {Tyagi}, {Surya}, {Nayak}, \& {Tripathi}}]{Banerjee-24}
{Banerjee}, B., {Narang}, M., {Manoj}, P., {et~al.} 2024,
  \bibinfo{title}{{Host-star Properties of Hot, Warm, and Cold Jupiters in the
  Solar Neighborhood from Gaia Data Release 3: Clues to Formation Pathways},}
  The Astronomical Journal, 168, 7, \dodoi{10.3847/1538-3881/ad429f}

\bibitem[{B. {Bitsch} {et~al.}(2013){Bitsch}, {Crida}, {Libert}, \&
  {Lega}}]{Bitsch-13}
{Bitsch}, B., {Crida}, A., {Libert}, A.~S., \& {Lega}, E. 2013,
  \bibinfo{title}{{Highly inclined and eccentric massive planets. I.
  Planet-disc interactions},} Astronomy \& Astrophysics, 555, A124,
  \dodoi{10.1051/0004-6361/201220310}

\bibitem[{A.~P. {Boss}(1997){Boss}}]{Boss-97}
{Boss}, A.~P. 1997, \bibinfo{title}{{Giant planet formation by gravitational
  instability.},} Science, 276, 1836, \dodoi{10.1126/science.276.5320.1836}

\bibitem[{A.~P. {Boss}(2024){Boss}}]{Boss-24}
{Boss}, A.~P. 2024, \bibinfo{title}{{Formation of Giant Planets by Gas Disk
  Gravitational Instability on Wide Orbits around Protostars with Varied
  Masses. II. Quadrupled Spatial Resolution and Beta Cooling},} \apj, 969, 157,
  \dodoi{10.3847/1538-4357/ad4ed4}

\bibitem[{L.~A. {Buchhave} {et~al.}(2018){Buchhave}, {Bitsch}, {Johansen},
  {Latham}, {Bizzarro}, {Bieryla}, \& {Kipping}}]{Buchhave-18}
{Buchhave}, L.~A., {Bitsch}, B., {Johansen}, A., {et~al.} 2018,
  \bibinfo{title}{{Jupiter Analogs Orbit Stars with an Average Metallicity
  Close to That of the Sun},} The Astrophysical Journal, 856, 37,
  \dodoi{10.3847/1538-4357/aaafca}

\bibitem[{L.~A. {Buchhave} {et~al.}(2012){Buchhave}, {Latham}, {Johansen},
  {Bizzarro}, {Torres}, {Rowe}, {Batalha}, {Borucki}, {Brugamyer}, {Caldwell},
  {Bryson}, {Ciardi}, {Cochran}, {Endl}, {Esquerdo}, {Ford}, {Geary},
  {Gilliland}, {Hansen}, {Isaacson}, {Laird}, {Lucas}, {Marcy}, {Morse},
  {Robertson}, {Shporer}, {Stefanik}, {Still}, \& {Quinn}}]{Buchhave-12}
{Buchhave}, L.~A., {Latham}, D.~W., {Johansen}, A., {et~al.} 2012,
  \bibinfo{title}{{An abundance of small exoplanets around stars with a wide
  range of metallicities},} Nature, 486, 375, \dodoi{10.1038/nature11121}

\bibitem[{L.~A. {Buchhave} {et~al.}(2014){Buchhave}, {Bizzarro}, {Latham},
  {Sasselov}, {Cochran}, {Endl}, {Isaacson}, {Juncher}, \&
  {Marcy}}]{Buchhave-14}
{Buchhave}, L.~A., {Bizzarro}, M., {Latham}, D.~W., {et~al.} 2014,
  \bibinfo{title}{{Three regimes of extrasolar planet radius inferred from host
  star metallicities},} Nature, 509, 593, \dodoi{10.1038/nature13254}

\bibitem[{M. {Deal} {et~al.}(2018){Deal}, {Alecian}, {Lebreton}, {Goupil},
  {Marques}, {LeBlanc}, {Morel}, \& {Pichon}}]{Deal-18}
{Deal}, M., {Alecian}, G., {Lebreton}, Y., {et~al.} 2018,
  \bibinfo{title}{{Impacts of radiative accelerations on solar-like oscillating
  main-sequence stars},} Astronomy \& Astrophysics, 618, A10,
  \dodoi{10.1051/0004-6361/201833361}

\bibitem[{E. {Delgado Mena} {et~al.}(2021){Delgado Mena}, {Adibekyan},
  {Santos}, {Tsantaki}, {Gonz{\'a}lez Hern{\'a}ndez}, {Sousa}, \& {Bertr{\'a}n
  de Lis}}]{Delgado-21}
{Delgado Mena}, E., {Adibekyan}, V., {Santos}, N.~C., {et~al.} 2021,
  \bibinfo{title}{{Chemical abundances of 1111 FGK stars from the HARPS GTO
  planet search program. IV. Carbon and C/O ratios for Galactic stellar
  populations and planet hosts},} \aap, 655, A99,
  \dodoi{10.1051/0004-6361/202141588}

\bibitem[{D.~A. {Fischer} \& J. {Valenti}(2005){Fischer} \&
  {Valenti}}]{Fischer-05}
{Fischer}, D.~A., \& {Valenti}, J. 2005, \bibinfo{title}{{The
  Planet-Metallicity Correlation},} The Astrophysical Journal, 622, 1102,
  \dodoi{10.1086/428383}

\bibitem[{L. {Ghezzi} {et~al.}(2021){Ghezzi}, {Martinez}, {Wilson}, {Cunha},
  {Smith}, \& {Majewski}}]{Ghezzi-21}
{Ghezzi}, L., {Martinez}, C.~F., {Wilson}, R.~F., {et~al.} 2021,
  \bibinfo{title}{{A Spectroscopic Analysis of the California-Kepler Survey
  Sample. II. Correlations of Stellar Metallicities with Planetary
  Architectures},} \apj, 920, 19, \dodoi{10.3847/1538-4357/ac14c3}

\bibitem[{G. {Gonzalez}(1997){Gonzalez}}]{Gonzalez-97}
{Gonzalez}, G. 1997, \bibinfo{title}{{The stellar metallicity-giant planet
  connection},} Monthly Notices of the Royal Astronomical Society, 285, 403,
  \dodoi{10.1093/mnras/285.2.403}

\bibitem[{L. {Grossman}(1972){Grossman}}]{Grossman-72}
{Grossman}, L. 1972, \bibinfo{title}{{Condensation in the primitive solar
  nebula},} \gca, 36, 597, \dodoi{10.1016/0016-7037(72)90078-6}

\bibitem[{Y. {Hasegawa} \& R.~E. {Pudritz}(2014){Hasegawa} \&
  {Pudritz}}]{Hasegawa-14}
{Hasegawa}, Y., \& {Pudritz}, R.~E. 2014, \bibinfo{title}{{Planet Traps and
  Planetary Cores: Origins of the Planet-Metallicity Correlation},} The
  Astrophysical Journal, 794, 25, \dodoi{10.1088/0004-637X/794/1/25}

\bibitem[{R. {Helled} {et~al.}(2014){Helled}, {Bodenheimer}, {Podolak},
  {Boley}, {Meru}, {Nayakshin}, {Fortney}, {Mayer}, {Alibert}, \&
  {Boss}}]{Helled-14}
{Helled}, R., {Bodenheimer}, P., {Podolak}, M., {et~al.} 2014,
  \bibinfo{title}{{Giant Planet Formation, Evolution, and Internal Structure},}
  in Protostars and Planets VI, ed. H.~{Beuther}, R.~S. {Klessen}, C.~P.
  {Dullemond}, \& T.~{Henning}, 643--665,
  \dodoi{10.2458/azu_uapress_9780816531240-ch028}

\bibitem[{N.~R. {Hinkel} {et~al.}(2014){Hinkel}, {Timmes}, {Young}, {Pagano},
  \& {Turnbull}}]{Hinkel-14}
{Hinkel}, N.~R., {Timmes}, F.~X., {Young}, P.~A., {Pagano}, M.~D., \&
  {Turnbull}, M.~C. 2014, \bibinfo{title}{{Stellar Abundances in the Solar
  Neighborhood: The Hypatia Catalog},} The Astronomical Journal, 148, 54,
  \dodoi{10.1088/0004-6256/148/3/54}

\bibitem[{J.~D. Hunter(2007)Hunter}]{Hunter-07}
Hunter, J.~D. 2007, \bibinfo{title}{Matplotlib: A 2D graphics environment,}
  Computing in Science \& Engineering, 9, 90, \dodoi{10.1109/MCSE.2007.55}

\bibitem[{S. {Ida} \& D.~N.~C. {Lin}(2004){Ida} \& {Lin}}]{Ida-04}
{Ida}, S., \& {Lin}, D.~N.~C. 2004, \bibinfo{title}{{Toward a Deterministic
  Model of Planetary Formation. II. The Formation and Retention of Gas Giant
  Planets around Stars with a Range of Metallicities},} The Astrophysical
  Journal, 616, 567, \dodoi{10.1086/424830}

\bibitem[{T. {Kutra} {et~al.}(2021){Kutra}, {Wu}, \& {Qian}}]{Kutra-21}
{Kutra}, T., {Wu}, Y., \& {Qian}, Y. 2021, \bibinfo{title}{{Super-Earths and
  Sub-Neptunes Are Insensitive to Stellar Metallicity},} \aj, 162, 69,
  \dodoi{10.3847/1538-3881/ac0431}

\bibitem[{J. {Maldonado} {et~al.}(2019){Maldonado}, {Villaver}, {Eiroa}, \&
  {Micela}}]{Maldonado-19}
{Maldonado}, J., {Villaver}, E., {Eiroa}, C., \& {Micela}, G. 2019,
  \bibinfo{title}{{Connecting substellar and stellar formation: the role of the
  host star's metallicity},} Astronomy \& Astrophysics, 624, A94,
  \dodoi{10.1051/0004-6361/201833827}

\bibitem[{C. {Marois} {et~al.}(2010){Marois}, {Zuckerman}, {Konopacky},
  {Macintosh}, \& {Barman}}]{Marois-10}
{Marois}, C., {Zuckerman}, B., {Konopacky}, Q.~M., {Macintosh}, B., \&
  {Barman}, T. 2010, \bibinfo{title}{{Images of a fourth planet orbiting HR
  8799},} Nature, 468, 1080, \dodoi{10.1038/nature09684}

\bibitem[{R. {Matsukoba} {et~al.}(2023){Matsukoba}, {Vorobyov}, {Hosokawa}, \&
  {Guedel}}]{Matsukoba-23}
{Matsukoba}, R., {Vorobyov}, E.~I., {Hosokawa}, T., \& {Guedel}, M. 2023,
  \bibinfo{title}{{Formation of a wide-orbit giant planet in a gravitationally
  unstable subsolar-metallicity protoplanetary disc},} Monthly Notices of the
  Royal Astronomical Society, 526, 3933, \dodoi{10.1093/mnras/stad3003}

\bibitem[{M. {Mayor} \& D. {Queloz}(1995){Mayor} \& {Queloz}}]{Mayor-95}
{Mayor}, M., \& {Queloz}, D. 1995, \bibinfo{title}{{A Jupiter-mass companion to
  a solar-type star},} Nature, 378, 355, \dodoi{10.1038/378355a0}

\bibitem[{M. {Mayor} {et~al.}(2003){Mayor}, {Pepe}, {Queloz}, {Bouchy},
  {Rupprecht}, {Lo Curto}, {Avila}, {Benz}, {Bertaux}, {Bonfils}, {Dall},
  {Dekker}, {Delabre}, {Eckert}, {Fleury}, {Gilliotte}, {Gojak}, {Guzman},
  {Kohler}, {Lizon}, {Longinotti}, {Lovis}, {Megevand}, {Pasquini}, {Reyes},
  {Sivan}, {Sosnowska}, {Soto}, {Udry}, {van Kesteren}, {Weber}, \&
  {Weilenmann}}]{Mayor-03}
{Mayor}, M., {Pepe}, F., {Queloz}, D., {et~al.} 2003, \bibinfo{title}{{Setting
  New Standards with HARPS},} The Messenger, 114, 20

\bibitem[{W. McKinney {et~al.}(2010)McKinney {et~al.}}]{mckinney2010data}
McKinney, W., {et~al.} 2010, \bibinfo{title}{Data structures for statistical
  computing in python,} in Proceedings of the 9th Python in Science Conference,
  Vol. 445, Austin, TX, 51--56

\bibitem[{C. {Mordasini} {et~al.}(2012){Mordasini}, {Alibert}, {Benz}, {Klahr},
  \& {Henning}}]{Mordasini-12}
{Mordasini}, C., {Alibert}, Y., {Benz}, W., {Klahr}, H., \& {Henning}, T. 2012,
  \bibinfo{title}{{Extrasolar planet population synthesis . IV. Correlations
  with disk metallicity, mass, and lifetime},} Astronomy \& Astrophysics, 541,
  A97, \dodoi{10.1051/0004-6361/201117350}

\bibitem[{G.~D. {Mulders}(2018){Mulders}}]{Mulders-18}
{Mulders}, G.~D. 2018, \bibinfo{title}{{Planet Populations as a Function of
  Stellar Properties},} in Handbook of Exoplanets, ed. H.~J. {Deeg} \& J.~A.
  {Belmonte} (Springer), 153, \dodoi{10.1007/978-3-319-55333-7_153}

\bibitem[{G.~D. {Mulders} {et~al.}(2016){Mulders}, {Pascucci}, {Apai},
  {Frasca}, \& {Molenda-{\.Z}akowicz}}]{Mulders-16}
{Mulders}, G.~D., {Pascucci}, I., {Apai}, D., {Frasca}, A., \&
  {Molenda-{\.Z}akowicz}, J. 2016, \bibinfo{title}{{A Super-solar Metallicity
  for Stars with Hot Rocky Exoplanets},} \aj, 152, 187,
  \dodoi{10.3847/0004-6256/152/6/187}

\bibitem[{M. {Narang} {et~al.}(2018){Narang}, {Manoj}, {Furlan}, {Mordasini},
  {Henning}, {Mathew}, {Banyal}, \& {Sivarani}}]{Narang-18}
{Narang}, M., {Manoj}, P., {Furlan}, E., {et~al.} 2018,
  \bibinfo{title}{{Properties and Occurrence Rates for Kepler Exoplanet
  Candidates as a Function of Host Star Metallicity from the DR25 Catalog},}
  The Astronomical Journal, 156, 221, \dodoi{10.3847/1538-3881/aae391}

\bibitem[{ {NASA Exoplanet Science Institute}(2020){NASA Exoplanet Science
  Institute}}]{nea}
{NASA Exoplanet Science Institute}. 2020, Planetary Systems Table, Last
  Accessed: 2024-12-18 IPAC, \dodoi{10.26133/NEA12}

\bibitem[{S. {Nayakshin}(2017){Nayakshin}}]{Nayakshin-17}
{Nayakshin}, S. 2017, \bibinfo{title}{{Dawes Review 7: The Tidal Downsizing
  Hypothesis of Planet Formation},} \pasa, 34, e002,
  \dodoi{10.1017/pasa.2016.55}

\bibitem[{M. {Nguyen} \& V. {Adibekyan}(2024){Nguyen} \&
  {Adibekyan}}]{Nguyen-24}
{Nguyen}, M., \& {Adibekyan}, V. 2024, \bibinfo{title}{{On the formation of
  super-Jupiters: core accretion or gravitational instability?},} \apss, 369,
  122, \dodoi{10.1007/s10509-024-04388-2}

\bibitem[{E.~A. {Petigura} {et~al.}(2018){Petigura}, {Marcy}, {Winn}, {Weiss},
  {Fulton}, {Howard}, {Sinukoff}, {Isaacson}, {Morton}, \&
  {Johnson}}]{Petigura-18}
{Petigura}, E.~A., {Marcy}, G.~W., {Winn}, J.~N., {et~al.} 2018,
  \bibinfo{title}{{The California-Kepler Survey. IV. Metal-rich Stars Host a
  Greater Diversity of Planets},} \aj, 155, 89,
  \dodoi{10.3847/1538-3881/aaa54c}

\bibitem[{B. {Ratcliffe} {et~al.}(2023){Ratcliffe}, {Minchev}, {Anders},
  {Khoperskov}, {Guiglion}, {Buck}, {Cunha}, {Queiroz}, {Nitschelm},
  {Meszaros}, {Steinmetz}, {de Jong}, {Nepal}, {Lane}, \&
  {Sobeck}}]{Ratcliffe-23}
{Ratcliffe}, B., {Minchev}, I., {Anders}, F., {et~al.} 2023,
  \bibinfo{title}{{Unveiling the time evolution of chemical abundances across
  the Milky Way disc with APOGEE},} Monthly Notices of the Royal Astronomical
  Society, 525, 2208, \dodoi{10.1093/mnras/stad1573}

\bibitem[{N.~C. {Santos} {et~al.}(2004){Santos}, {Israelian}, \&
  {Mayor}}]{Santos-04}
{Santos}, N.~C., {Israelian}, G., \& {Mayor}, M. 2004,
  \bibinfo{title}{{Spectroscopic [Fe/H] for 98 extra-solar planet-host stars.
  Exploring the probability of planet formation},} Astronomy \& Astrophysics,
  415, 1153, \dodoi{10.1051/0004-6361:20034469}

\bibitem[{N.~C. {Santos} {et~al.}(2011){Santos}, {Mayor}, {Bonfils},
  {Dumusque}, {Bouchy}, {Figueira}, {Lovis}, {Melo}, {Pepe}, {Queloz},
  {S{\'e}gransan}, {Sousa}, \& {Udry}}]{Santos-11}
{Santos}, N.~C., {Mayor}, M., {Bonfils}, X., {et~al.} 2011,
  \bibinfo{title}{{The HARPS search for southern extrasolar planets. XXV.
  Results from the metal-poor sample},} \aap, 526, A112,
  \dodoi{10.1051/0004-6361/201015494}

\bibitem[{N.~C. {Santos} {et~al.}(2017{\natexlab{a}}){Santos}, {Adibekyan},
  {Figueira}, {Andreasen}, {Barros}, {Delgado-Mena}, {Demangeon}, {Faria},
  {Oshagh}, {Sousa}, {Viana}, \& {Ferreira}}]{Santos-17}
{Santos}, N.~C., {Adibekyan}, V., {Figueira}, P., {et~al.} 2017{\natexlab{a}},
  \bibinfo{title}{{Observational evidence for two distinct giant planet
  populations},} Astronomy \& Astrophysics, 603, A30,
  \dodoi{10.1051/0004-6361/201730761}

\bibitem[{N.~C. {Santos} {et~al.}(2017{\natexlab{b}}){Santos}, {Adibekyan},
  {Dorn}, {Mordasini}, {Noack}, {Barros}, {Delgado-Mena}, {Demangeon}, {Faria},
  {Israelian}, \& {Sousa}}]{Santos-17b}
{Santos}, N.~C., {Adibekyan}, V., {Dorn}, C., {et~al.} 2017{\natexlab{b}},
  \bibinfo{title}{{Constraining planet structure and composition from stellar
  chemistry: trends in different stellar populations},} Astronomy \&
  Astrophysics, 608, A94, \dodoi{10.1051/0004-6361/201731359}

\bibitem[{S. {Savvidou} \& B. {Bitsch}(2023){Savvidou} \&
  {Bitsch}}]{Savvidou-23}
{Savvidou}, S., \& {Bitsch}, B. 2023, \bibinfo{title}{{How to make giant
  planets via pebble accretion},} \aap, 679, A42,
  \dodoi{10.1051/0004-6361/202245793}

\bibitem[{S. Seabold \& J. Perktold(2010)Seabold \& Perktold}]{Seabold-10}
Seabold, S., \& Perktold, J. 2010, \bibinfo{title}{statsmodels: Econometric and
  statistical modeling with python,} in 9th Python in Science Conference

\bibitem[{S.~G. {Sousa} {et~al.}(2011){Sousa}, {Santos}, {Israelian}, {Mayor},
  \& {Udry}}]{Sousa-11}
{Sousa}, S.~G., {Santos}, N.~C., {Israelian}, G., {Mayor}, M., \& {Udry}, S.
  2011, \bibinfo{title}{{Spectroscopic stellar parameters for 582 FGK stars in
  the HARPS volume-limited sample. Revising the metallicity-planet
  correlation},} Astronomy \& Astrophysics, 533, A141,
  \dodoi{10.1051/0004-6361/201117699}

\bibitem[{D.~S. {Spiegel} {et~al.}(2011){Spiegel}, {Burrows}, \&
  {Milsom}}]{Spiegel-11}
{Spiegel}, D.~S., {Burrows}, A., \& {Milsom}, J.~A. 2011, \bibinfo{title}{{The
  Deuterium-burning Mass Limit for Brown Dwarfs and Giant Planets},} The
  Astrophysical Journal, 727, 57, \dodoi{10.1088/0004-637X/727/1/57}

\bibitem[{H.-Y. {Teng} {et~al.}(2023){Teng}, {Sato}, {Kuzuhara}, {Takarada},
  {Omiya}, {Harakawa}, {Izumiura}, {Kambe}, {Yilmaz}, {Bikmaev}, {Selam},
  {Brandt}, {Xiao}, {Yoshida}, {Itoh}, {Ando}, {Kokubo}, \& {Ida}}]{Teng-23}
{Teng}, H.-Y., {Sato}, B., {Kuzuhara}, M., {et~al.} 2023,
  \bibinfo{title}{{Revisiting planetary systems in the Okayama Planet Search
  Program: A new long-period planet, RV astrometry joint analysis, and a
  multiplicity-metallicity trend around evolved stars},} Publications of the
  Astronomical Society of Japan, 75, 1030, \dodoi{10.1093/pasj/psad056}

\bibitem[{V. {Van Eylen} {et~al.}(2018){Van Eylen}, {Agentoft}, {Lundkvist},
  {Kjeldsen}, {Owen}, {Fulton}, {Petigura}, \& {Snellen}}]{VanEylen-18}
{Van Eylen}, V., {Agentoft}, C., {Lundkvist}, M.~S., {et~al.} 2018,
  \bibinfo{title}{{An asteroseismic view of the radius valley: stripped cores,
  not born rocky},} Monthly Notices of the Royal Astronomical Society, 479,
  4786, \dodoi{10.1093/mnras/sty1783}

\bibitem[{P. Virtanen {et~al.}(2020)Virtanen, Gommers, Oliphant, Haberland,
  Reddy, Cournapeau, Burovski, Peterson, Weckesser, Bright, {van der Walt},
  Brett, Wilson, Millman, Mayorov, Nelson, Jones, Kern, Larson, Carey, Polat,
  Feng, Moore, {VanderPlas}, Laxalde, Perktold, Cimrman, Henriksen, Quintero,
  Harris, Archibald, Ribeiro, Pedregosa, {van Mulbregt}, \& {SciPy 1.0
  Contributors}}]{Virtanen-20}
Virtanen, P., Gommers, R., Oliphant, T.~E., {et~al.} 2020,
  \bibinfo{title}{{{SciPy} 1.0: Fundamental Algorithms for Scientific Computing
  in Python},} Nature Methods, 17, 261, \dodoi{10.1038/s41592-019-0686-2}

\bibitem[{E. {Vorobyov} \& C. {Schoenhacker}(2025){Vorobyov} \&
  {Schoenhacker}}]{Vorobyov-25}
{Vorobyov}, E., \& {Schoenhacker}, C. 2025, \bibinfo{title}{{Protoplanet and
  Proto-Brown Dwarf Clumps in Gravitationally Unstable Protoplanetary Disks of
  Various Metallicity},} Universe, 11, 116, \dodoi{10.3390/universe11040116}

\bibitem[{M.~L. Waskom(2021)Waskom}]{Waskom2021}
Waskom, M.~L. 2021, \bibinfo{title}{seaborn: statistical data visualization,}
  Journal of Open Source Software, 6, 3021, \dodoi{10.21105/joss.03021}

\bibitem[{A. {Weeks} {et~al.}(2025){Weeks}, {Van Eylen}, {Huber}, {Kawata},
  {Stokholm}, {Aguirre B{\o}rsen-Koch}, {Pinilla}, {R{\o}rsted}, {Winther}, \&
  {Berger}}]{Weeks-25}
{Weeks}, A., {Van Eylen}, V., {Huber}, D., {et~al.} 2025, \bibinfo{title}{{A
  link between rocky planet composition and stellar age},} Monthly Notices of
  the Royal Astronomical Society, 539, 405, \dodoi{10.1093/mnras/staf474}

\bibitem[{L.~M. {Weiss} {et~al.}(2018){Weiss}, {Isaacson}, {Marcy}, {Howard},
  {Petigura}, {Fulton}, {Winn}, {Hirsch}, {Sinukoff}, {Rowe}, \& {California
  Kepler Survey}}]{Weiss-18}
{Weiss}, L.~M., {Isaacson}, H.~T., {Marcy}, G.~W., {et~al.} 2018,
  \bibinfo{title}{{The California-Kepler Survey. VI. Kepler Multis and Singles
  Have Similar Planet and Stellar Properties Indicating a Common Origin},} \aj,
  156, 254, \dodoi{10.3847/1538-3881/aae70a}

\bibitem[{R.~F. {Wilson} {et~al.}(2022){Wilson}, {Ca{\~n}as}, {Majewski},
  {Cunha}, {Smith}, {Bender}, {Mahadevan}, {Fleming}, {Teske}, {Ghezzi},
  {J{\"o}nsson}, {Beaton}, {Hasselquist}, {Stassun}, {Nitschelm},
  {Garc{\'\i}a-Hern{\'a}ndez}, {Hayes}, \& {Tayar}}]{Wilson-22}
{Wilson}, R.~F., {Ca{\~n}as}, C.~I., {Majewski}, S.~R., {et~al.} 2022,
  \bibinfo{title}{{The Influence of 10 Unique Chemical Elements in Shaping the
  Distribution of Kepler Planets},} \aj, 163, 128,
  \dodoi{10.3847/1538-3881/ac3a06}

\bibitem[{J. {Wimarsson} {et~al.}(2020){Wimarsson}, {Liu}, \&
  {Ogihara}}]{Wimarsson-20}
{Wimarsson}, J., {Liu}, B., \& {Ogihara}, M. 2020, \bibinfo{title}{{Promoted
  mass growth of multiple, distant giant planets through pebble accretion and
  planet-planet collision},} Monthly Notices of the Royal Astronomical Society,
  496, 3314, \dodoi{10.1093/mnras/staa1708}

\bibitem[{J.~K. {Zink} {et~al.}(2023){Zink}, {Hardegree-Ullman},
  {Christiansen}, {Petigura}, {Boley}, {Bhure}, {Rice}, {Yee}, {Isaacson},
  {Fernandes}, {Howard}, {Blunt}, {Lubin}, {Chontos}, {Pidhorodetska}, \&
  {MacDougall}}]{Zink-23}
{Zink}, J.~K., {Hardegree-Ullman}, K.~K., {Christiansen}, J.~L., {et~al.} 2023,
  \bibinfo{title}{{Scaling K2. VI. Reduced Small-planet Occurrence in
  High-galactic-amplitude Stars},} \aj, 165, 262,
  \dodoi{10.3847/1538-3881/acd24c}

\end{thebibliography}
\bibliographystyle{aasjournalv7}



\end{document}